\documentclass{aa}

\usepackage{graphicx}
\usepackage{txfonts}
\usepackage{amsmath}
\usepackage{amssymb}
\usepackage{wasysym}
\usepackage{color}
\usepackage{xcolor}
\usepackage{epsfig}
\usepackage{longtable}
\usepackage{pdflscape}
\usepackage{mathtools}
\usepackage{float}
\usepackage{footnote}
\usepackage{epstopdf}
\usepackage{units}
\usepackage{enumerate}
\usepackage{pdflscape}

\usepackage{txfonts}
\usepackage{subfiles} % Best loaded last in the preamble

\begin{document}

   \title{Structural parameters of 389 local Open Clusters 
   \thanks{The tables with cluster members and mean cluster parameters are only available in electronic form at the CDS via anonymous ftp to cdsarc.u-strasbg.fr (130.79.128.5) 
or via http://cdsarc.u-strasbg.fr/viz-bin/qcat?J/A+A/?/?}}

   \author{Y. Tarricq\inst{1},
          C. Soubiran\inst{1},
          L. Casamiquela\inst{1},
          A. Castro-Ginard\inst{2, 3}, 
          J. Olivares\inst{1, 4}, 
          N. Miret-Roig\inst{1, 5}, 
          P. A. B. Galli\inst{1}}
   \authorrunning{Y. Tarricq et al.}
   \institute{Laboratoire  d’Astrophysique  de  Bordeaux,  Univ.  Bordeaux,  CNRS,  B18N,  allée  Geoffroy  Saint-Hilaire,  33615  Pessac,  France\\
              \email{yoann.tarricq@u-bordeaux.fr}
         \and   
             Institut de Ciències del Cosmos, Universitat de Barcelona (IEEC-UB), Martí i Franquès 1, E-08028 Barcelona, Spain
         \and
             Leiden Observatory, Leiden University, Niels Bohrweg 2, 2333 CA Leiden, Netherlands.
         \and
             Instituto de Astrofísica de Canarias, E-38205 La Laguna, Tenerife, Spain; Universidad de La Laguna, Dpto. Astrofísica, E-38206 La Laguna, Tenerife, Spain.
         \and
             University of Vienna, Department of Astrophysics, Türkenschanzstraße 17, 1180 Wien, Austria
             }

   \date{\today}

% \abstract{}{}{}{}{} 
% 5 {} token are mandatory
 
  \abstract
   {The distribution of member stars in the surroundings of an open cluster can shed light on the process of its formation, evolution and dissolution. The analysis of structural parameters of open clusters as a function of their age and position in the Galaxy brings constraints on theoretical models of cluster evolution. The \emph{ Gaia} catalogue is very appropriate to find members of open clusters at large distance from their centers. }
   {We aim at revisiting the membership lists of open clusters from the solar vicinity, in particular by extending these membership lists to the peripheral areas thanks to \emph{Gaia} EDR3. We then take advantage of these new lists of members to study the morphological properties and the mass segregation levels of the clusters.}
   {We used the clustering algorithm HDBSCAN on \emph{Gaia} parallaxes and proper motions to systematically look for members up to 50 pc from the cluster centers. We fitted a King's function on the radial density profile of these clusters and a Gaussian Mixture Model on their two dimensional distribution of members to study their shape. We also evaluated the degree of mass segregation of the clusters and the correlations of these parameters with the age and Galactic position of the clusters.}
   {Our methodology performs well on 389 clusters out of the 467 selected ones, including several recently discovered clusters that were poorly studied until now. We report the detection of vast coronae around almost all the clusters and report the detection of 71 open clusters with tidal tails, multiplying by more than four the number of such structures identified. We find the size of the cores to be on average smaller for old clusters than for young ones. Also, the overall size of the clusters seems to slightly increase with age while the fraction of stars in the halo seems to decrease. As expected the mass segregation is more pronounced in the oldest clusters but a clear trend with age is not seen.}
  {Open Clusters are more extended than previously expected, regardless of their age. The decrease in the proportion of stars populating the clusters halos highlights the different cluster evaporation processes and the short timescales they need to affect the clusters. Reported parameters like cluster sizes or mass segregation levels all depend on cluster ages but can not be described as single functions of time.}

\keywords{Galaxy: kinematics and dynamics --
            Galaxy: structure --
            open clusters and associations: general--
            methods: statistical --
            Surveys: Gaia}

   \maketitle
%---------------------------------------------------------------
\section{Introduction}
Open clusters (OCs) are essential objects to better understand the evolution of the stellar disc of the Milky Way. Most stars of the disc are believed to be born in OCs \citep{lad03} which dissipate into the field due to relaxation-driven mass loss or tidal perturbations, as recently reviewed by \cite{kru19}. The morphology of OCs is directly related to these processes. OCs have first to survive an initial gas expulsion \citep{baumgardt2007} following the formation of their first stars. Then, they experience a violent phase of relaxation during which stars can be expelled and form tail-like structures depending on star formation efficiency \citep{dinn2020b} and the timescales of gas expulsion \citep{din2020}, among other processes. In addition, young OCs may keep in their morphology the imprint of substructures from their parent molecular clouds \citep{alves2020}. Recently, the hierarchical formation scenario has been proposed \citep{mcmi2007} in order to explain the evidence of mass segregation in young clusters such as the Orion Nebula Cluster \citep{hill1998}. Standard dynamical evolution being unable to explain such levels of mass segregation in young clusters \citep{bon1998}, this scenario postulates that stars form in small clumps which later merge to form larger mass segregated systems. On the other hand, older clusters are governed by both internal and external effects. Equipartition of kinetic energy via two-body relaxation has a direct consequence on the distribution of stars within a cluster. Massive stars within a cluster move towards its center whereas lighter stars move towards its outskirts in a mass segregation process \citep{mat1984,kro1995,delafuente1996}. At the same time, gravitational perturbations by giant molecular clouds, tidal stripping due to the galactic potential or spiral arm shocks, perturb the cohesion of star clusters and shape escaping stars into "S-shaped" tidal structures \citep{kup2008,din2020}. Eventually, at their final stages, clusters may disintegrate \citep{lam2006} while their members mix into the galactic field. The spatial distribution of members in OCs of various ages and in different environments can shed light on all these processes. In particular structural parameters such as the size of the core, the presence of a halo or a tidal tail in the peripheral region and the degree of mass segregation can bring new constraints to theoretical models.

The successive publication of the second \emph{Gaia} data release \citep[DR2,][]{gaiaDR2} and of the early data release 3 \citep[EDR3,][]{gaiaEDR3}, led to what could be called a revolution in the study of OCs. With almost 1.5 billion sources with a full astrometric solution (position, proper motions and parallaxes), the census of OCs as well as their characterization have been drastically improved. Many studies took advantage of \emph{Gaia} DR2 to compute new memberships or to detect new clusters, thanks to very different techniques and algorithms. \citet{can18} computed membership probabilities of 1229 OCs that were identified prior to \emph{Gaia} by \citet{dias2002} and \citet{khar2013} using the clustering algorithm UPMASK \citep{kro14}. Hundreds of new OCs and their members were identified by \citet{cas18, cas19, cas20} who developed a machine learning approach to spot over-densities in the five dimensional parameter space of positions, parallaxes and proper motions. These works released a catalogue of more than 600 new open clusters. \citet{sim19} visually inspected stellar distributions in the galactic coordinates and proper motion space and identified 207 new cluster candidates. Also \citet{liu2019} used a friend-of-friend method already widely used in the galaxy cluster community to identify 76 unreported clusters. \citet{kounkel20} applied the unsupervised machine learning algorithm HDBSCAN to identify not only clusters but also moving groups and associations within 1 kpc. More recently \citet{can20b}, hereafter CAN+20, published a catalog of 2017 OCs previously identified by the aforementioned authors and determined in a homogeneous way their memberships, distances and ages. Most of these large scale studies are focused on the inner parts of clusters and are therefore unable to provide members in the peripheral regions of OCs. 

The combination of the striking precision of the \emph{Gaia} astrometric measurements and its all sky coverage allowed the detection of prominent structures around some OCs by several groups. \citet{roe2019}, \citet{mei2019} and \citet{jer21} characterized the tidal tails of the Hyades at large spatial scale with different methods. The tidal tails of Coma Berenices, Ruprecht 147,  Praesepe, Blanco 1, NGC 2506 and NGC 752 were discovered successively by \citet{tang2019}, \citet{yeh2019}, \citet{roe2019}, \citet{zhang2020}, \citet{gao2020a} and \citet{bha2021}. \citet{mei21} studied ten nearby (located closer than 500 pc), prominent and young OCs and identified around almost all of them an extended population of stars, referred as a corona. In general, OC shapes can be described with a dense core and an outer halo (or corona) having a low density of stars \citep{art1964}. As pointed out by \citet{nit2002} and more recently by \citet{mei21}, halos are much more extended than the cores and they are suspected to comprise a large number of cluster members. 

The most complete and enlightening studies of the morphology of OCs are those conducted in 3D. However, their major drawback is that a 3D study requires to convert parallaxes into distances which is not a trivial transformation. As established by \citet{bai2015}, it requires the use of Bayesian inference and the choice of a prior, prior which depends on the aim of the study. Moreover, \emph{Gaia} parallaxes suffer from systematic errors and biases which have significantly improved in \emph{Gaia}-EDR3 compared to \emph{Gaia}-DR2 \citep{lin2021}, but still translate into an elongated shape of the clusters along the line of sight. Consequently 3D studies are limited to very nearby (<500pc) OCs \citep{pie2021}. In order to study the morphology of clusters further than 500 pc from the Sun it is more efficient to work in 2D.

In this paper, we perform a membership analysis using \emph{Gaia} EDR3 for a sample of known OCs closer than 1.5 kpc and older than 50 Myr, with a particular effort to detect new members at large distance from their center. Taking advantage of these new memberships, we study the shape of the OCs projected on the plane of the sky. We measure the core and tidal radii of the clusters, the elongation and the size of their halo, we look for tails and we quantify the level of mass segregation. We evaluate how these properties correlate with the age and the galactic position of the clusters.

This paper is organised as follows. Section \ref{sec:method} describes the selection of clusters, the \emph{Gaia} EDR3 query, the clustering algorithm and the new memberships. We analyse the radial profile in Sect. \ref{sec:RDP} and the different populations of  each cluster through Gaussian Mixture Models (GMM) in Sect. \ref{sec:GMM}. In Sect. \ref{sec:mass_segregation}, we present our study of the mass segregation and Sect. \ref{sec:conclusion} summarises our results. 
\section{Clustering}\label{sec:method}
\subsection{Data}\label{sec:data}

We selected all the OCs from CAN+20 closer than 1.5 kpc from the Sun and older than 50 Myr. These cuts were implemented after some tests of the adopted methodology, described in details in Sect \ref{sec:clustering_hdbscan}. We noticed lower performances for clusters younger than 50 Myr which are often embedded in their star forming region. On another hand clusters more distant than 1.5 kpc have larger astrometric errors making the membership analysis less reliable. We found the adopted cuts to be the best compromise to obtain reliable results on a large sample. This leaves us with 467 clusters. The CAN+20 catalogue includes the most recent improvements of the OC census based on \emph{Gaia} DR2, previously reported in \cite{can20b, can18} and \cite{ cas18, cas19, cas20}. Here we take advantage of the exquisite astrometric precision of \emph{Gaia} EDR3 to re-visit the memberships of the selected clusters on a wide area around their center. We used the mean proper motions, parallaxes and positions calculated by CAN+20 to query the \emph{Gaia} archive for each OC as follows: 
\begin{itemize}
    \item we queried cone of 50 pc radius around the center of each cluster,
    \item we used the cluster dispersion in proper motion from CAN+20 to perform cuts at 10 $\sigma$ in proper motion to discard very discrepant stars and help the clustering algorithm,
    \item we considered only stars with $G<18$ mag and with a Renormalized Unit Weight Error (RUWE) lower than 1.4, following the recommendation of \citet{edr3_catalogue}.
    \item for clusters closer than 500 pc, which span a very wide area on the sky, we applied an additional cut in parallax in order to limit the number of stars in the query. Based on $\varpi_{cluster}$, the mean parallax of CAN+20, we left a margin of 200 pc so that all the stars with parallaxes $\varpi$ verifying the following relation were selected: $1/\varpi_{cluster}-200 pc<1/\varpi<1/\varpi_{cluster}+200 pc$.
\end{itemize}

For each cluster, these cuts allowed us to discard a significant amount of stars having astrometric measurements inconsistent with the cluster's mean astrometric parameters.

\subsection{Clustering}\label{sec:clustering_hdbscan}

We used the clustering algorithm Hierarchical Density-Based Spatial Clustering of Applications with Noise (HDBSCAN) \citep{HDBSCAN} using its python implementation \citep{hdbscan_python} to perform our membership study. It aims at improving the performance of the widely used density-based algorithm DBSCAN \citep{DBSCAN}, which was successfully applied to the search of OCs by \cite{cas18, cas19, cas20}. One of the main advantages of HDBSCAN over DBSCAN is that it is able to detect overdensities of varying density in a dataset . To do so, HDBSCAN adds  a hierarchical approach to DBSCAN. To detect a cluster, DBSCAN draws hyperspheres of radii $\epsilon$ around each star and considers as a cluster the points inside a hypersphere containing more than \texttt{minPts}. In other words, a cluster for DBSCAN is defined as the points within an overdensity which is more populated than the chosen parameter \texttt{minPts}. We refer the reader to \cite{cas18} for a detailed explanation of DBSCAN. HDBSCAN does not depend on the radius $\epsilon$ of a hypersphere as it scans all values of $\epsilon$ and uses them to build a hierarchical tree merging these different results. Clusters are defined by two parameters, the parameter \texttt{min\_cluster\_size} which is equivalent to the parameter \texttt{minPts} of DBSCAN and the parameter \texttt{min\_samples} which sets how conservative the algorithm is. Higher values of \texttt{min\_samples} will discard the clusters with the lowest contrast with respect to the background and consider them as noise even if their number of members is higher than \texttt{min\_cluster\_size}. All clusters containing less stars than \texttt{min\_cluster\_size} will automatically be classified as noise. Once the hierarchical tree of the dataset has been built, HDBSCAN offers two options to select the clusters: the Excess of Mass (EoM) clustering and the leaf clustering. We selected the clusters according to the leaf method which chooses the clusters located at the lowest level of the tree. As noted by \citet{hunt21}, the leaf clustering method almost always performed better in the identification of OCs and we therefore adopted this method. 

We ran HDBSCAN on each dataset resulting from the \emph{Gaia} EDR3 query described in Sect. \ref{sec:data}. Following \cite{kounkel20} we chose as initial parameters \texttt{min\_cluster\_size}=40 and  \texttt{min\_samples}=25. We tested this analysis with different input parameters on a subset of clusters representative of our sample and found this choice as the best compromise. If HDBSCAN did not identify a cluster, we lowered iteratively these parameters and try again the clustering. We ran HDBSCAN on the three dimensions space of the parallax and proper motions, ($\mu_{\alpha^{*}}$, $\mu_{\delta}, \varpi$), but not on the sky coordinates to avoid penalizing the stars in the cluster outskirts\footnote{We do not use the sky coordinates to compute our membership lists like \cite{cas18, cas19, cas20} because we aim at studying the halos of known clusters}. For every cluster, we ran HDBSCAN 100 times, each time with a new sample of individual ($\mu_{\alpha^{*}}$, $\mu_{\delta}, \varpi$) randomly generated from their uncertainties taking into account the correlations between them, as done by \cite{can18}. This process allowed us to compute membership probabilities: the membership probability of a star corresponds to the frequency with which it was considered as a member by HDBSCAN. 

In some cases HDBSCAN identified several clusters, either statistical clusters, asterisms or other physical groups located in the same field. In this case, we took advantage of the previous information that we have for that particular cluster, and we only considered the group identified by HDBSCAN with mean proper motions and parallax the closest to the value computed by CAN+20. This is for instance the case of the cluster NGC~7063 which has two close neighbors: ASCC~113 and UPK~113. With this additional filter, we could separate the three clusters at each run. This constraint also allowed us to systematically discard statistical groups detected by our algorithm which do not correspond to the targeted cluster.

Visual inspection of the results (and of the CMD) obtained for the 467 clusters showed a successful membership list for 389 OCs. It performed poorly on some particular cases that we discarded: (1) the clusters with too few stars (less than 30), (2) the clusters with neighbours which overlap in the same field of the query and are also close in the parallax-proper motion space. The latter situation happened in particular around star forming regions and results from the large radius of our query. Even though we discarded most of the star forming regions by considering only clusters older than 50 Myr old, some of the youngest clusters of our sample are still close from their birth location. We also noticed during this visual inspection that the probability distribution of the members depends on the Galactic coordinates of the cluster and on the density of field stars surrounding the cluster. For the sake of clarity, we decided to use 0.5 as a probability cutoff for the membership list which we considered as the best compromise for the large variety of clusters in our sample. A table containing all the members with a membership probability higher than 0.1 is available in the CDS\footnote{The readers will therefore be able to adopt a different threshold that is more specific to their scientific objectives.}.

\subsection{New memberships}\label{sec:new_memberships}

\begin{figure}[h]
\centering
\includegraphics[width=\textwidth/2]{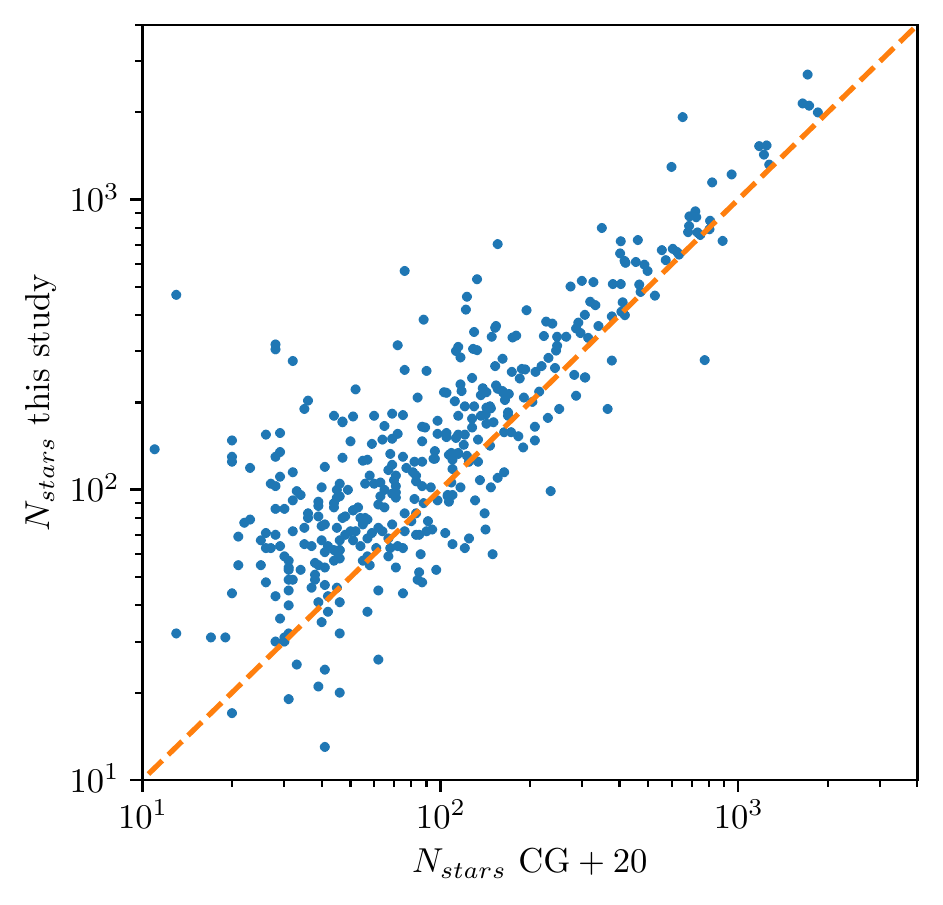}
\caption{Comparison between the number of stars in CAN+20 and in this study. The dashed line shows the identity relationship.}
\label{f:nbstars}
\end{figure}

For the majority of the 389 OCs, we find a number of members significantly larger than CAN+20 as seen in Fig. \ref{f:nbstars} (we recall that CAN+20 published the list of members with a membership probability higher than 0.7). A striking case is UBC~480 with 13 members in CAN+20 and 470 in our analysis. Our study increases the number of cluster members by a factor of 36. The number of members of NGC~6716 has also been increased by more then 850\%, rising from 70 to 568. NGC~6716 was identified by \citet{gri1990} and CAN+20 as a sparsely populated cluster but we found it as a quite populated cluster with a dense core and a large halo (or corona). 

Figure \ref{f:example} shows an example of the results of our clustering procedure on two well known clusters: Blanco~1 and NGC 2682. We can clearly see that in both cases, we recover almost all members identified by CAN+20 and that we extend the memberships way further than the cores, reaching the limits of our search radius. For NGC 2682 we identify many halo stars, confirming previous findings of \citet{car19}. In the case of Blanco~1, a tidal tail already reported by \citet{zhang2020} is also detected by our method in addition to halo stars.  We detect vast coronae around a significant number of clusters, similarly to \citet{mei21} who performed a 3D analysis on ten prominent and nearby clusters. We detect similar structures in the five clusters we have in common. These coronae extend until the edge of our search radius, suggesting that they are even larger. We detected these coronae even for distant clusters such as NGC~2477 located at 1415 pc. 

\begin{figure*}[h]
\centering
\includegraphics[width=\textwidth]{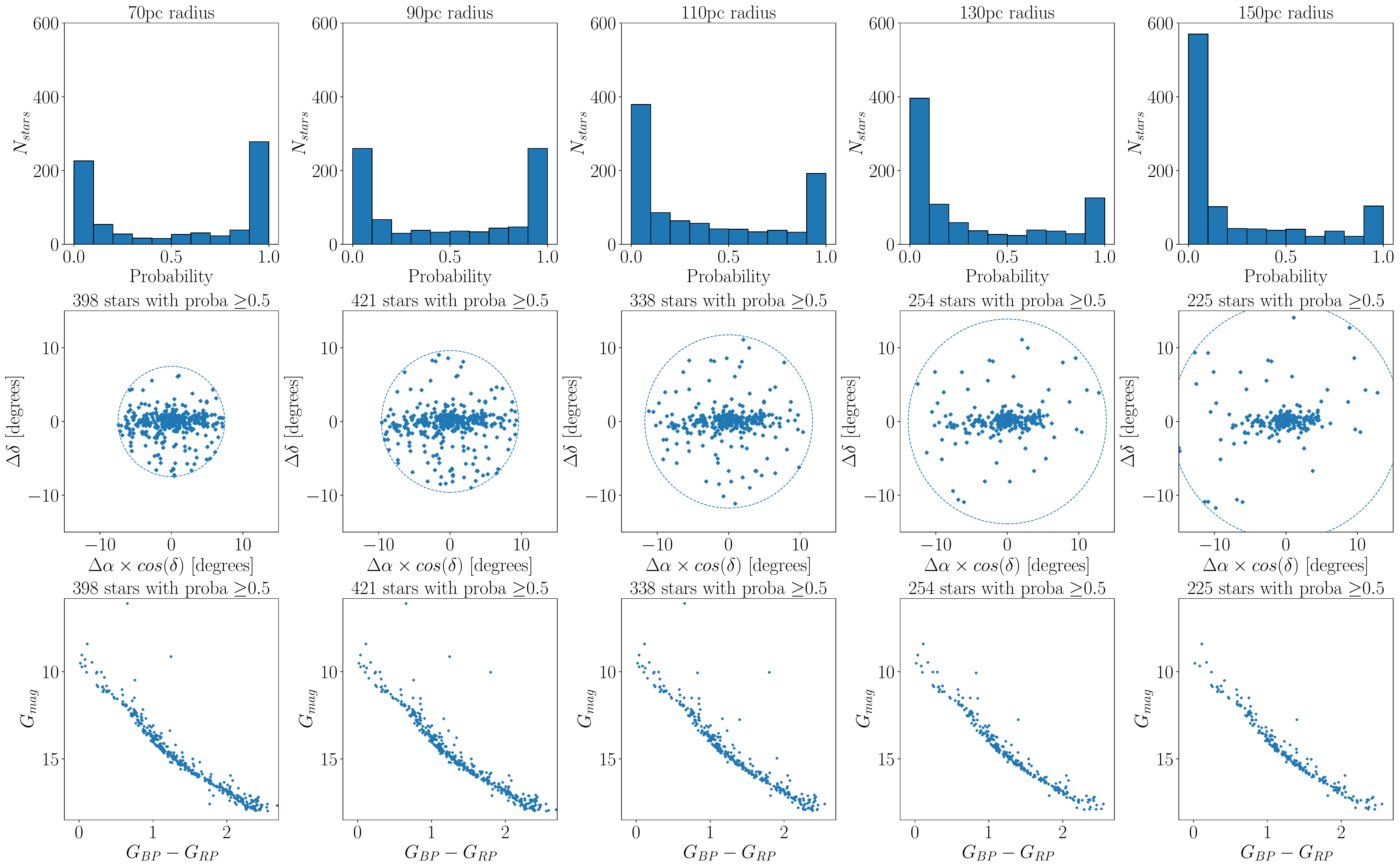}

\caption{Probability distribution (top row), distribution of the recovered members of COIN-Gaia 13 on the celestial sphere (middle row) and color magnitude diagram (bottom row) for concentric search radius going from 70 to 150 pc by steps of 20 pc. On each of the middle panel we show the edge of the search radius in degrees with the blue dotted circle and we indicated the number of stars passing our probability cutoff.}
\label{fig:coin_gaia_1}
\end{figure*}

In order to test the hypothesis that the coronae can extend to very large distances from the core and that recovered members are limited by the search radii, we inspected the results of our method on the recently discovered OC COIN-Gaia 13 for different search radii. We queried the \emph{Gaia} archive as described in Sect. \ref{sec:data} but we did 5 concentric cone searches of increasing radius with steps of 20 pc. We show the resulting probability distributions and members recovered up to a radius of 150 pc on Fig. \ref{fig:coin_gaia_1}. We can see that we keep identifying members until the edge of the cone search even at a radius of 150 pc. However, the number of members stopped increasing after reaching a radius of 80 pc which we interpret as the limit of the corona. This also shows that for very extended fields, a probability cutoff of 0.5 might not be the best compromise anymore.  

\begin{figure*}[h]
\centering
\includegraphics[width=\textwidth]{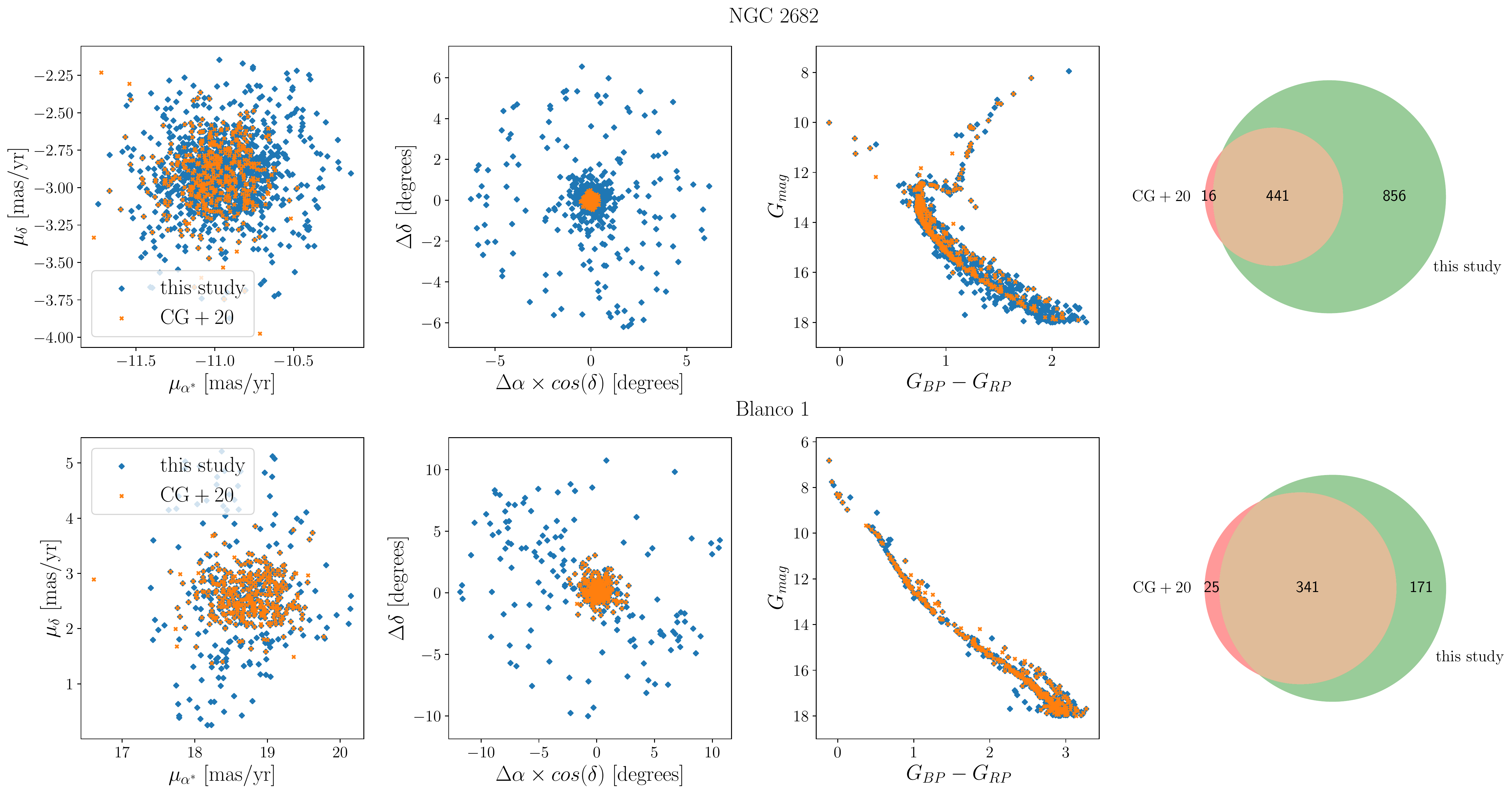}
\caption{Example of the results of our clustering procedure on NGC~2682 (upper panels) and Blanco~1 (lower panels). For each cluster, the three scatter plots represent (from left to right) a comparison between the members from CAN+20 (in orange) and ours (in blue) in the proper motion space, in the equatorial coordinate space, and in the color magnitude diagram. The rightmost panel shows for both clusters a Venn diagram comparing the number of members in both studies with members from CAN+20 in red, our members in green and the overlap in orange.}
\label{f:example}
\end{figure*}

\begin{figure*}[h]
\centering
\includegraphics[width=\textwidth]{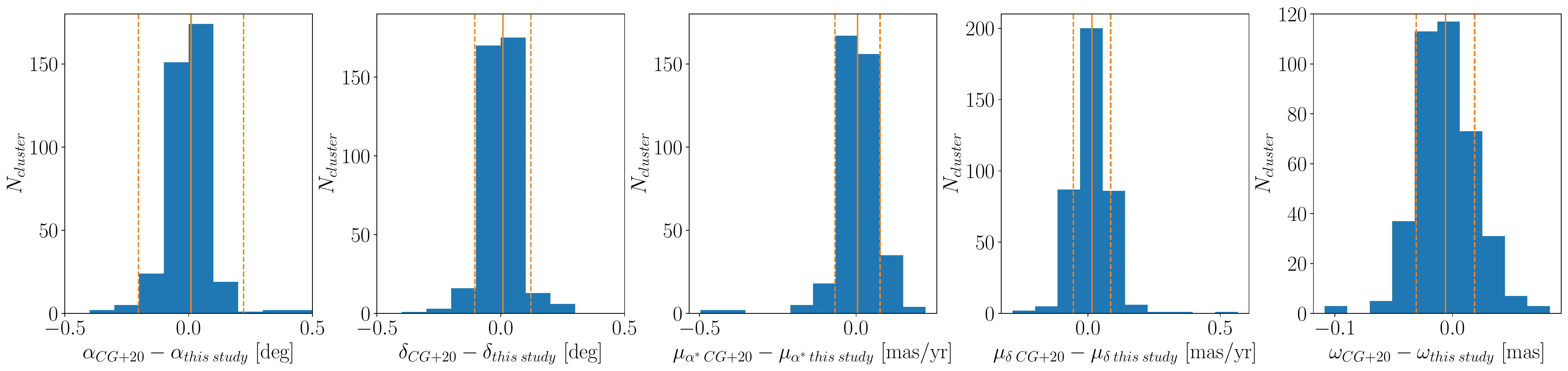}
\caption{Distribution of the residuals of the mean cluster parameters of CAN+20 and the ones calculated in this study. The orange solid line represents the mean of the distribution and the orange dashed lines show the 1 $\sigma$ standard deviation to the mean. For clarity, the offset between the mean positions calculated here and the previous known mean positions are only shown in the range (-0.5, 0.5) degrees, even if nine and five OCs lie beyond this limit for respectively the right ascension and the declination. }
\label{f:comparison_mean_params}
\end{figure*}

We computed the mean position and parallax of each OC by finding the maximum density point through a kernel density estimation. The mean proper motions are computed differently. The proper motion distributions are too flat to properly assume the maximum density point. Therefore we used the same method as CAN+20: we calculated the median value after removing outliers away from the median by more than three median absolute deviations (MAD). The mean astrometric parameters of our OC sample and the comparison to those of CAN+20 are presented in Fig. \ref{f:comparison_mean_params}. The mean of the residuals of the comparison to CAN+20 is well centered on zero for both the positions and proper motions. Inevitably, for some clusters the members of CAN+20 and ours are different: either some members were not retrieved by HDBSCAN, or we have many more members now (which represent the majority of cases). This creates significant differences in the mean centers and proper motions of some clusters compared to CAN+20, especially for the low populated ones. However for parallaxes, the distribution of the residuals shows a negligible offset of -0.008 mas and a MAD of 0.015 mas showing the agreement between our values and those of CAN+20.

\section{Radial Density Profiles}\label{sec:RDP}
We aim at measuring the structural parameters of the OCs in our sample, based on our new lists of members extended to the outskirts of the clusters. The radial density profile (RDP) is a good indicator to study the extension of the spatial distribution of the clusters members. Once obtained, the resulting density profile can be characterized by means of the fit of the widely used King empirical function \citep{King62}.

\subsection{Fitting procedure}\label{fitting}

The King's profile is widely used to fit the radial density profile of OCs although it was first introduced to describe the surface density of globular clusters \citep{King62}. It is defined as: 
\begin{align}
\begin{aligned}
n(R)=\left\{
\begin{array}{ll}
k \cdot \left( \frac{1}{\sqrt{1+(R/R_c)^{2}}} - \frac{1}{\sqrt{1+(R_t/R_c)^{2}}} \right)^{2}+c & \mbox{if } R<R_t \\
c & \mbox{if } R \ge R_t,
\end{array}\right.
\end{aligned}\label{eq:king}
\end{align}
where $k$ is a scaling constant related to the central density, $R_c$ the core radius, $R_t$ the tidal radius and $n(R)$ is the surface density in stars per squared parsecs. Following \citet{kup2010}, we added a constant $c$ to the original formula of \citet{King62}, also in stars per squared parsecs. We expect $c$ to be close to zero since we are considering the most reliable cluster
members. This constant improved significantly the quality of the fits for many clusters. The core radius is defined as the radius for which the value of the density is equal to half the central density. The tidal radius is the radius where the cluster becomes indistinguishable from the field \citep{King62}. In our case the tidal radius is therefore the radius for which the density is equal to $c$. 

The first step in order to fit a model such as the King profile to a cluster is to determine its radial density profile. To do so, we first needed to calculate the distance between each cluster's stars and its center. The centers of the clusters are computed as described in Sect \ref{sec:method}. Some of the clusters of our sample such as Ruprecht~98 have high declinations and the distribution of their members in the sky is therefore subject to strong projection effects. Some clusters, especially the most nearby ones, are also sensitive to projection effects due to the curvature of the celestial sphere, especially for nearby clusters. To avoid these biases, we projected each stars coordinates on the plane of the sky, tangential to the celestial sphere at the coordinates of the clusters centers, as suggested by \citet{vdw06} and \citet{oli18}. The projected coordinates are defined for each cluster star as: 

\begin{align}
\begin{aligned}
x &= D \cdot \sin(\alpha-\alpha_c) \cdot \cos(\delta) \\
y &= D \cdot \cos(\delta_c) \cdot \sin(\delta) - \sin(\delta_c) \cdot \cos(\delta) \cdot \cos(\alpha-\alpha_c),
\end{aligned}\label{eq:projected_coordinated}
\end{align}
where D is the heliocentric distance of the cluster computed by CAN+20 in pc, and $\alpha_c$ and $\delta_c$ are the clusters mean right ascension and declination.

The radial distance R of each star to the center of the cluster is: 
\begin{align}
\begin{aligned}
R &= \sqrt{x^2+y^2}.
\end{aligned}\label{eq:radial_distcance}
\end{align}

We divided the spatial distribution of the stars on these projected coordinates in concentric rings. We used 10 bins of one parsec width for the inner parts of the clusters and then we progressively increased the width of these rings. We computed the density defined as the number of stars per square parsec in each ring.

\begin{figure}[h]
\centering
\includegraphics[width=\textwidth/2]{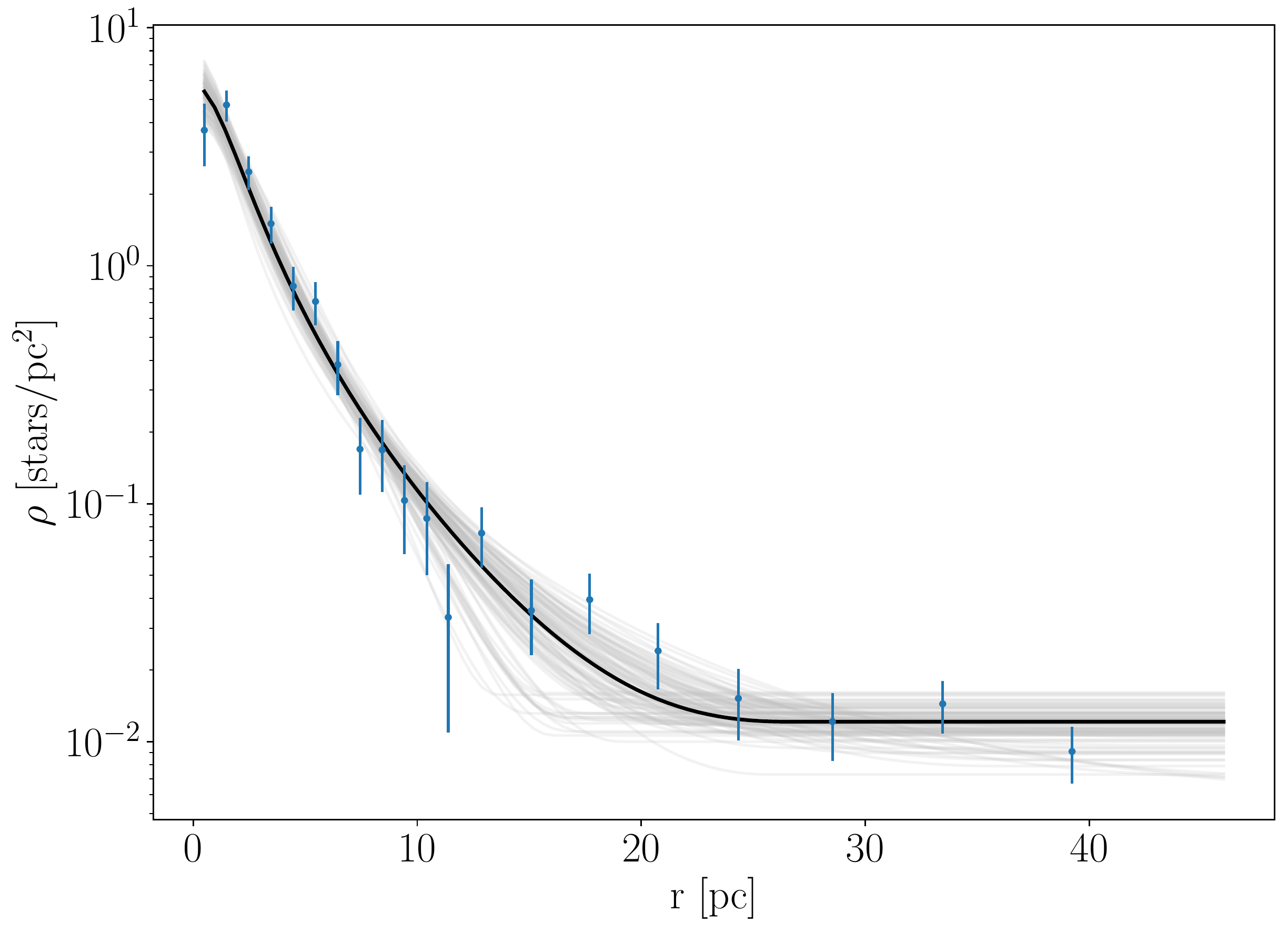}
\includegraphics[width=\textwidth/2]{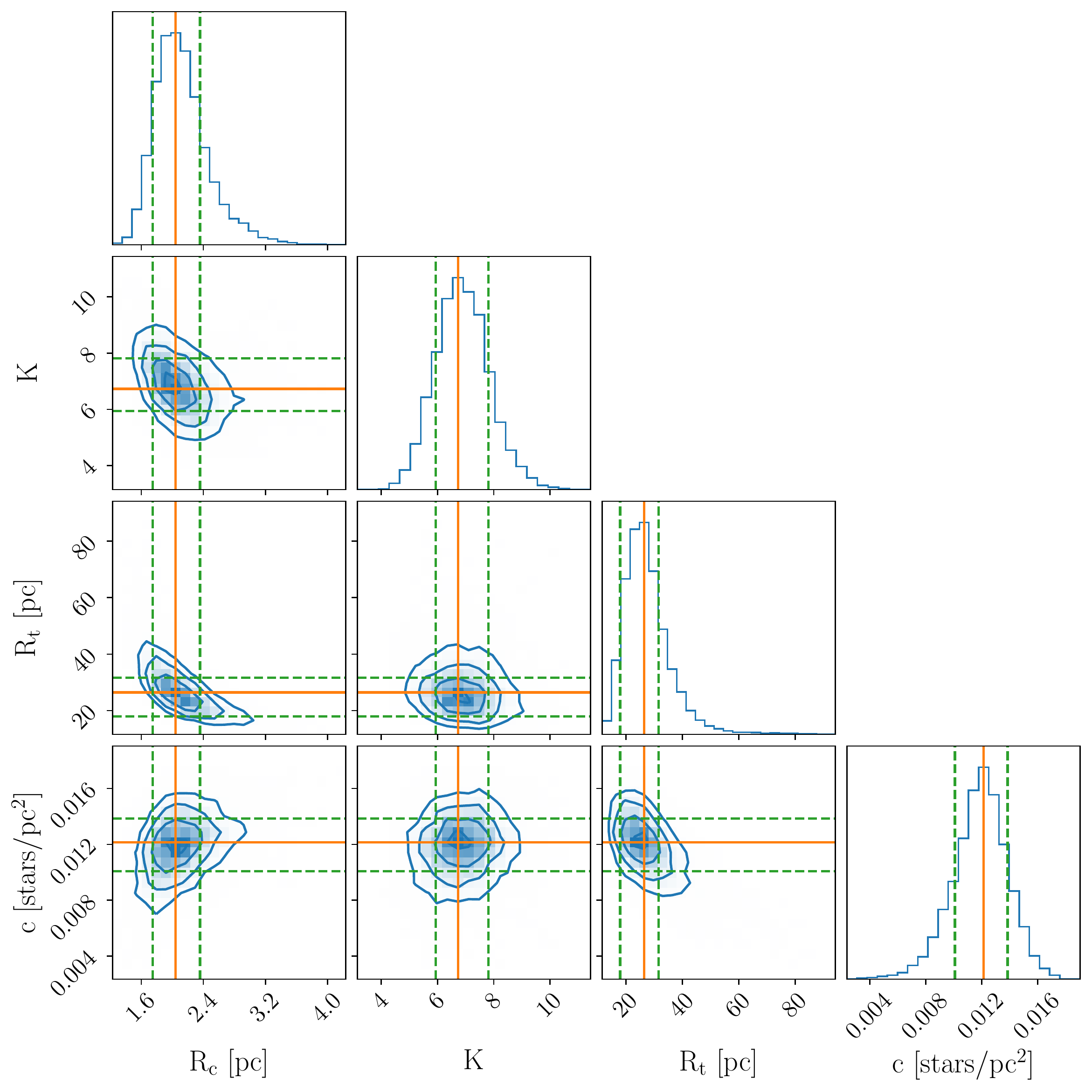}

\caption{Results of the King's profile fit on the cluster NGC~752. In the top panel, the blue dots are shown with Poissonian uncertainties. It also shows the best fit obtained with a ML estimator (black solid line), defined as the mode of the distributions of the parameters obtained though the 64 000 fits performed. The grey lines represent the uncertainties on the fits: we show 100 fits taken from the posterior distribution of the ML. The bottom panel shows the corresponding projection of the parameters posterior distribution. The orange lines show the mode of each distribution and the green dashed line shows the 68\% highest density interval (HDI).}
\label{fig:ngc752}
\end{figure}

We fitted the King's profile with a Maximum Likelihood (ML) estimator considering Poissonian uncertainties for each point. We used the MCMC sampler \texttt{emcee} \citep{emcee} and initialized eight "walkers" (two per parameter to fit). For each walker we assigned 10,000 iterations to converge, and we discarded the first 2,000 iterations to compute the posterior. As recommended, the convergence of the chains was systematically checked based on the integrated auto-correlation time \citep{goo2020}. The results of our fitting procedure are shown as an example for the cluster NGC~752 in Fig. \ref{fig:ngc752} where we found a core radius $R_c=2.04^{+0.31}_{-0.3}$ pc and a tidal radius of $R_t=26.45^{+5.15}_{-8.46}$. We applied the aforementioned procedure on the 233 clusters in our sample counting more than 100 members. We chose this lower limit in order to have sufficient number of stars in the circular rings. 

We only considered as satisfactory results the fits for which no flag was risen by the integrated auto-correlation time regarding the convergence of the chains of the fitting procedure. We also discarded the determinations of $R_c$ with errors greater than 2.5 pc and the determinations of $R_t$ with errors greater than 15 pc. This leaves us with estimations of $R_c$ and $R_t$ for respectively 164 and 145 OCs. Both of these quality cuts are mostly useful to discard the cases where the tidal radii are poorly constrained. Indeed, because of the sparse nature of OCs and, in some cases, of the contamination by field stars, the determination of the tidal radii is more challenging than the determination of the core radii \citep[p.~356]{reviewOCsfriel}. 

\subsection{Discussion}

The tidal radius estimated in this study is a parameter of the King's radial density profile and shall not be confused with the Jacobi radii introduced by \citet[p.~450]{binney_tremaine}. The Jacobi radius $R_j$ is often referred to as the tidal radius but is only a crude estimate of it. Unlike the tidal radius, the Jacobi radius does not involve the density to be equal to zero at $R=R_j$. 

\begin{figure}[h]
\centering
\includegraphics[width=\textwidth/2]{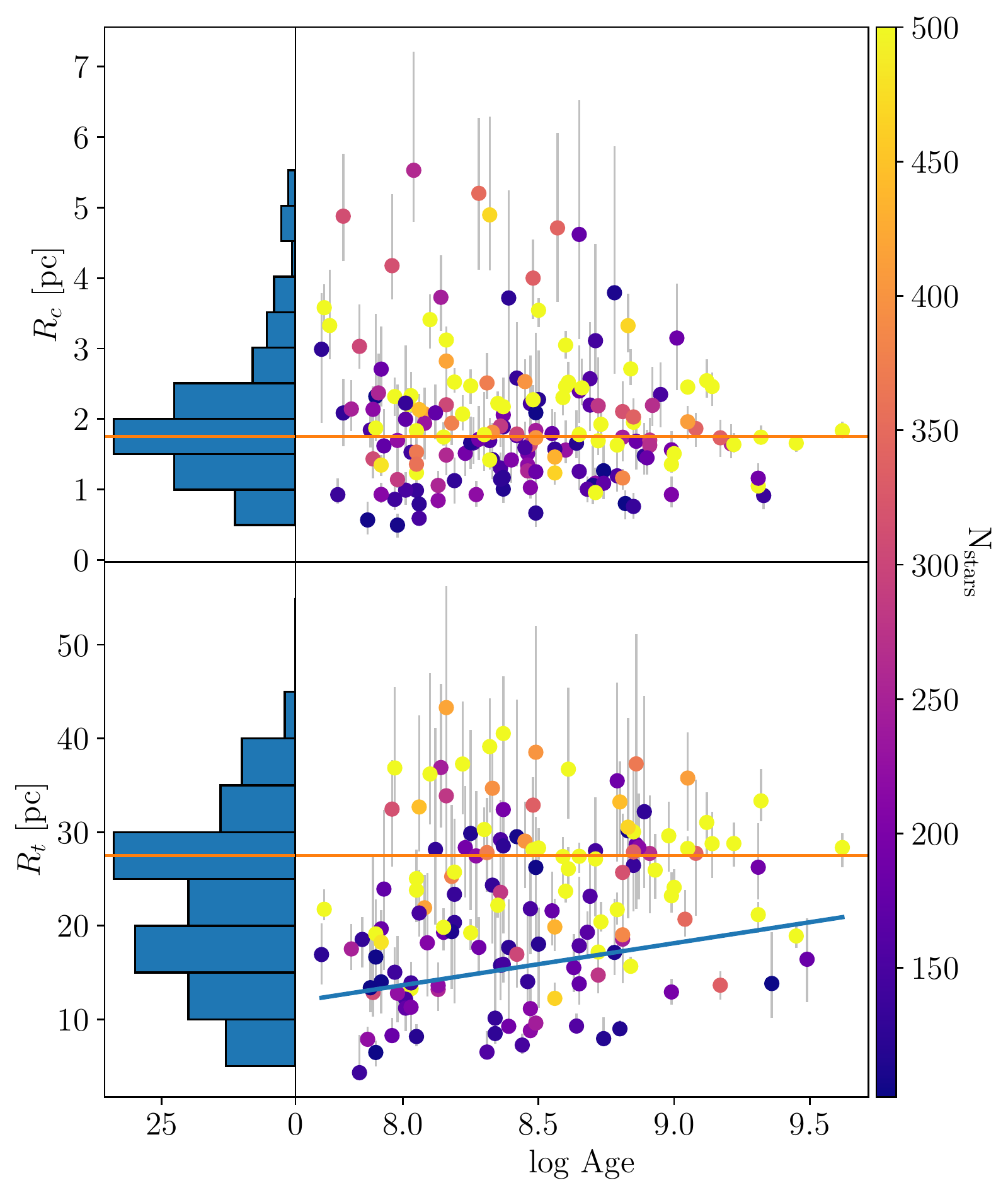}
\caption{Fitted core radii $R_c$ (top) and tidal radii $R_t$ (bottom) shown as a function of the logarithm of the cluster ages and their corresponding histograms. The color bar stands for the number of cluster members and the mode of the distributions is overplotted with the orange solid line. A linear regression of the tidal radii versus age has been fitted with a least square method (blue line).}
\label{fig:rt_rc_distribution}
\end{figure}

The fitted cores and tidal radii for each cluster are shown as a function of their ages in Fig. \ref{fig:rt_rc_distribution}. The core and tidal radii are computed as the mode of the parameter distributions of our ML procedure's chains. The uncertainties represent the lower and upper bound of the 68\% HDI of the ML chains. The vast majority of clusters have a core radius between 1 and 2.5 pc, regardless of their age and number of members. The most frequent value of the core radius is $\sim$1.85 pc. We can also note that the vast majority of the clusters with less than 250 members (in blue in the figure) have values of $R_c$ slightly smaller than the mode of the distribution while more populated clusters tend to have larger values. Finally, the dispersion of the core radius decreases with the increasing age of the clusters. This indicates that even if young clusters can have very concentrated cores, this feature is more common for old clusters. This is in agreement with the hypothesis discussed by \citet{heggie_hut_2003} that the evolution of the inner parts of the cluster is dominated by two-body relaxation causing the cluster core to shrink. Two-body relaxation is also known to cause mass segregation in clusters: massive stars concentrate in the cores of the clusters while less massive stars move in their outskirts. It could be connected to the observed decrease of the core radius with age. As more massive stars concentrate towards the cores of the clusters, the cores' gravitational potential increases which causes it to be denser. Mass segregation will be studied in details in Sect. \ref{sec:mass_segregation}. We want to point out that this decrease in the dispersion of the core radii could also be explained statistically as the clusters which deviates the most from the mode of the distribution are the ones with the largest errors. 

Looking at the bottom panel of Fig. \ref{fig:rt_rc_distribution}, we can see that the distribution of the tidal radius is bimodal. It peaks around 29 pc with a secondary peak at $\sim$18 pc. The majority of clusters with less than 250 members (in blue) have values of $R_t\sim$18 pc or lower while almost all of the populated clusters (in yellow) have larger values. Additionally, the tidal radius has a mild increase with the age of the clusters. We illustrate this increase by overplotting a linear regression of the tidal radii versus the age of the clusters. To perform the fit, we used a simple least square method taking into account the uncertainties in the tidal radii. We obtained values of 3.82$\pm$1.52 and -17.48$\pm$12.95 respectively for the slope and the y-intercept of the fit. This increase could again be connected to mass segregation or to cluster evaporation: because more stars have moved to the outskirts of the older clusters they are more likely to be torn off from the clusters. Consequently, this could produce an increase of the tidal radius with age.

\begin{figure}[h]
\centering
\includegraphics[width=\textwidth/2]{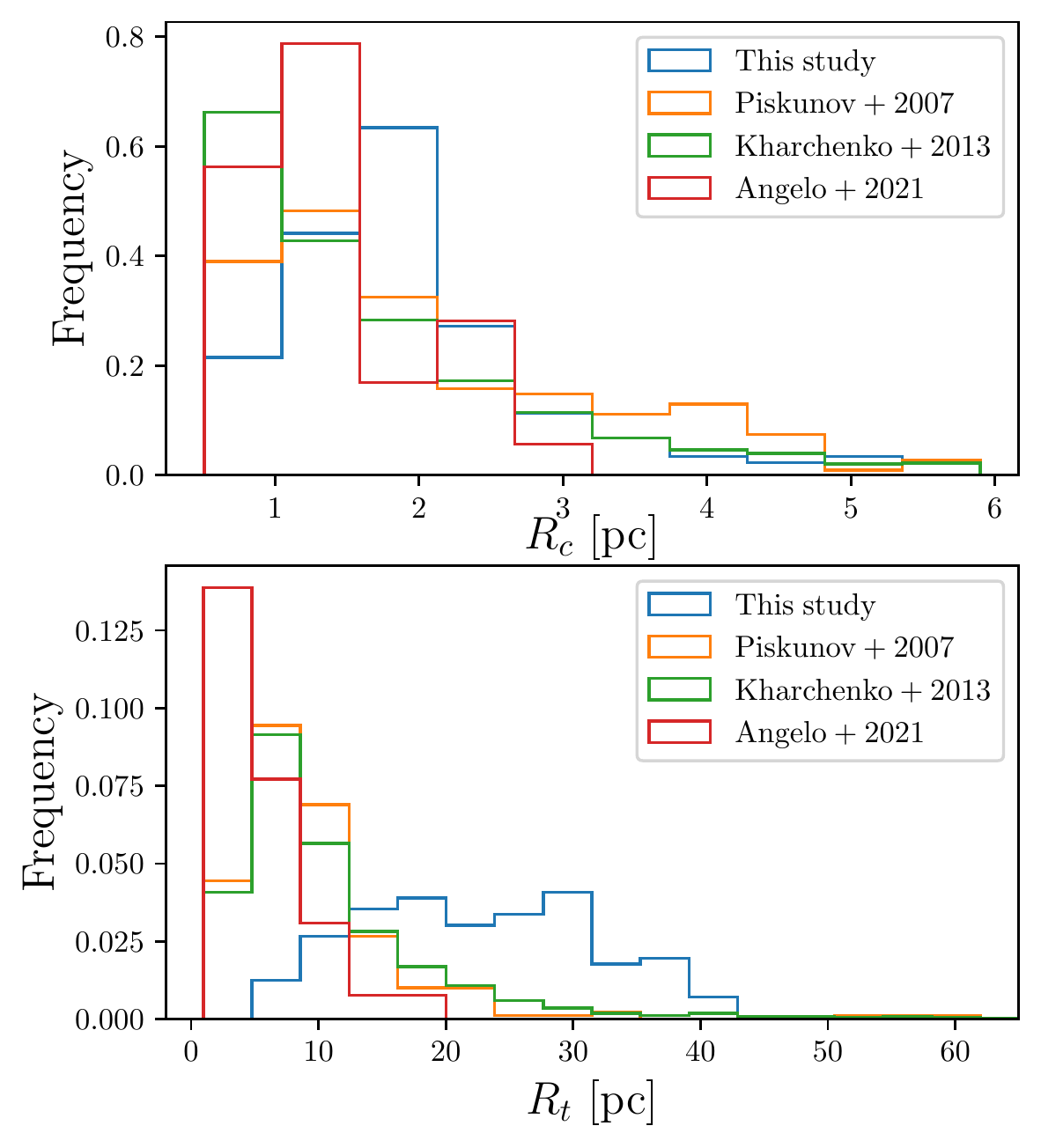}
\caption{Distribution of the core (top) and tidal (bottom) radii computed in this study and by \citet{pisk2007}, \citet{khar2013} and \citet{ang21}.}
\label{fig:comparison_rc_rt}
\end{figure}

Core and tidal radii of OCs have often been determined in the past through the fit of a King profile.  The most extensive catalog of radii was published before the launch of the \emph{Gaia} mission by \citet{khar2013}. Also before \emph{Gaia}, \citet{pisk2007} published a catalog of radii for 236 OCs out of the 650 clusters with reliable memberships from the catalogue ASCC-2.5. More recently, \citet{ang21} studied in details the structural parameters of 38 OCs with \emph{Gaia} DR2 data. The comparison of our determinations of core and tidal radii with these three studies is shown in Fig. \ref{fig:comparison_rc_rt}. We note that \citet{ang21} looked for members at a maximum radius of $1^{\circ}$ around the clusters centers computed by \citet{dias2002}. This is equivalent for most of their clusters to a radius smaller than 30 pc. We therefore have very different list of members, which makes a close comparison difficult. Nevertheless, we can compare the distributions of $R_c$ and $R_t$ as shown in Fig. \ref{fig:comparison_rc_rt}, the similarity between the distribution of $R_c$ in all the studies is striking while the tidal radii computed here are much larger than those computed in the other studies. This is a direct consequence of our choice to search for members at larger distances from the center of each cluster, compared to the previous studies. On another hand, in the case of M~67/NGC~2682, \citet{car19} looked for members up to 150 pc around the center of the cluster. They fitted a King function to the radial density profile and estimated a tidal radius of 80 pc while we find a value of $31.08^{+1.37}_{-1.48}$ pc. This suggests, as previously reported by \citet{oli18} that the determination of the tidal radius is highly dependant of the size of the survey. Therefore our distribution of tidal radii is likely truncated due to our queries around each cluster limited to 50 pc. 

Because we consider only cluster members in our fitting procedure, the $c$ constant from Eq. \ref{eq:king} (which is equivalent to a field constant density) should be close to 0. When it is higher, it gives us an estimation of the number of contaminants for each cluster. The median proportion of contamination in our sample is $\sim$13\%. This estimation of the contamination rate is biased for some clusters, leading to a high contamination rate. As explained in Sect. \ref{sec:new_memberships}, our distribution of members is likely truncated for populated clusters. This leads to an underestimation of the tidal radius of these clusters with typical values of $\sim$30 pc. The members detected beyond this estimation of the tidal radius act as a background field density in the fit, leading to an overestimation of the $c$ constant i.e. of the contamination rate. Moreover, the King profile might not be the best way to describe the density of some clusters with extended halos \citep{kup2010}, especially if they present an elongation like for Blanco~1: tidal tails stars will also act here as a background field density. That is why we fitted a GMM on the spatial distribution of members (see Sect. \ref{sec:GMM}).

\citet{sch2006} noted an increase of the size of the clusters with their altitude above the galactic plane and that this increase was especially significant for clusters older than $\sim$22~Myr. They also reported that large clusters were found at large Galactocentric distances. They concluded that clusters located within the solar orbit and with a low inclination of their orbit with respect to the Galactic plane are likely to be rapidly dissolved by encounters with giant molecular clouds or by Galactic tidal stripping. On the contrary, clusters with an orbit outside of the solar one and reaching high altitudes are more likely to survive longer and to not be stripped of their members. We do not see any of those correlations, even when we divide our sample of OCs in different age bins so we can not confirm these findings. More recently, \cite{dib2018} also failed to confirm these findings.
\section{Gaussian Mixture Models}\label{sec:GMM}
The function described in the previous section assumes a circular distribution of the members. In order to study the morphology of the OCs without making this assumption, we fitted a GMM on the spatial distribution of members of each OC. A GMM is a probabilistic model assuming that the data can be described by a combination of a finite number of Gaussian distributions.

\subsection{Fitting procedure}\label{sec:GMM_method}

To get rid of projection effets due to clusters located at high galactic latitudes and to members located far from the clusters' centers, we used Eq. \ref{eq:projected_coordinated} to project the galactic latitude and longitude of each star in a cluster on a plane tangential to the celestial sphere. In order to fit a GMM on these coordinates, we used a variational GMM with a Dirichlet process prior\footnote{Algorithm implemented in the scikit-learn python package \citep{sklearn}}. This algorithm is a variant of the classical GMM and allows to infer the effective number of components from the data. Usually, classical GMM fitting take advantage of Expectation-Maximization (EM) algorithm. In the EM algorithm, the parameters of the Gaussians are randomly initialized (the user can also provide a first guess) and the algorithm computes the probability of each data point to belong to each component. The parameters of each Gaussian are tuned in order to maximise the likelihood of the data under this model. Tuning the parameters of each Gaussian over a sufficient number N of iterations always allows to converge to a local maximum of the likelihood. The variational inference extends the EM approach by adding information through a prior distribution: the Dirichlet process. With the Dirichlet process prior, the number of components set by the user is only used as an upper bound, the algorithm will automatically draw the number of components from the data, activating a component and attributing stars to this component only if necessary: if the maximum number of components is set to 3 but the Dirichlet process only detects 2, it will set the relative weight of one of the component to $\sim$0. 

As explained in Sect. \ref{sec:method}, we realised by visual inspection that some clusters presented an elongation in their outskirts which could correspond to a tidal tail. A fraction of these OCs has a number of members comprised between 50 and 100. So in order to characterize as many tidal tails as possible, we fitted a GMM on all the clusters with more than 50 members. We also noticed that most of the clusters present a prominent core and an extended halo. Consequently we systematically tried to fit three components to the sky distribution of the members of each cluster. One component would correspond to the core of the cluster, the second one to the eventual tidal tail and the third one to the cluster's halo or coronae. Because not all the clusters show a tidal tail, the Dirichlet process is therefore well suited for our purpose: if only two components are detected, the weight of the component standing for the tidal tail is supposed to be close to zero. 

Since there is a stochastic initialisation in the variational inference algorithm, the fit does not always converge towards the same solution. Therefore, in order to have a better estimation of the parameters of the Gaussians, we run the fitting algorithm for each cluster a thousand times. The parameters of the resulting Gaussians were then chosen as the mode of the resulting distributions and their standard errors were computed as : $MAD/\sqrt{N_{iter}}$ with $N_{iter}=1000$. In addition we noticed the fits were better when the algorithm is forcing the gaussians to be concentric. 

\begin{figure}[t]
\centering
\includegraphics[width=\textwidth/2]{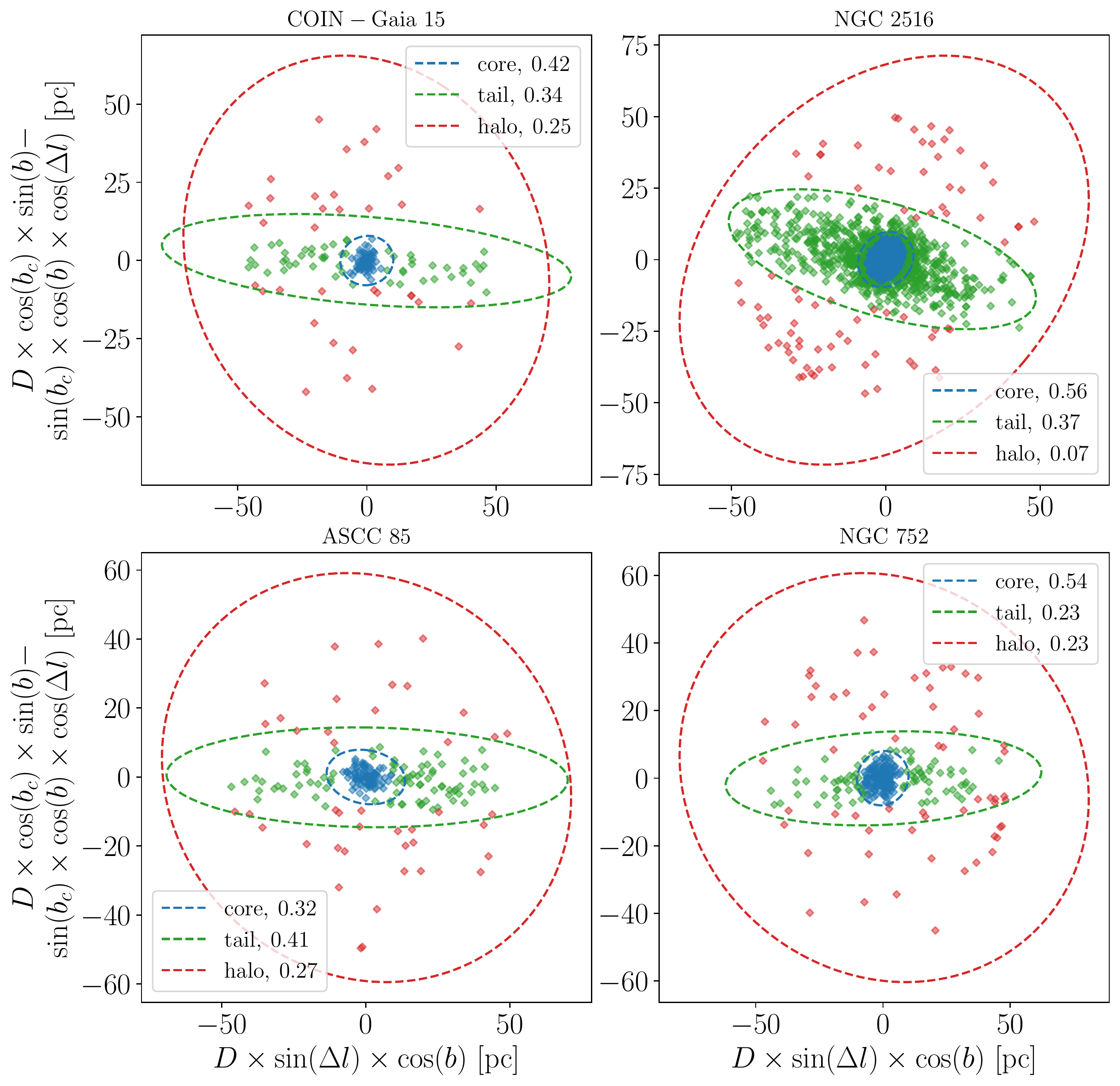}
\caption{Example of four clusters for which we detect a tidal tail. The blue, green and red ellipses represent the 3 $\sigma$ ellipse fitted on the distribution of the stars standing respectively for the core, the tidal tail and the halo. The stars are colored according to which components they are the most likely to belong to. The relative weights of each component is indicated on each panel.}
\label{f:tidal_tails}
\end{figure}

For some clusters, the algorithm found a prominent and elongated component of the GMM which could possibly be associated with a tidal tail. However we could not find a clear cut in weights or in eccentricity to separate clusters with and without elongation, owing to the large variety of clusters and environments we are dealing with. In particular, the most populated clusters (more than 500 members) with a dense core always end up with a second component having a significant weight even if it does not represent a tidal tail but more likely the outskirts of the core. In most of these cases the results with two components were sufficient to fit the distribution . We also tried to separate the clusters with and without a tidal tail using the length of the semi-major axis of the second component, the number of stars attributed to each components or the ratio between the semi minor and the semi major axis of the second component but we did not find an ideal way to separate the two subsamples with accuracy. That is why we visually identified them. We defined a subsample of 71 OCs with this feature. We show  the results for four clusters which present a remarkable elongation in  Fig. \ref{f:tidal_tails}. A prominent halo is also noticeable around the cores of the four clusters. The legend in each panel of Fig. \ref{f:tidal_tails} shows the weights attributed to each component by the algorithm. 

In the following, for the 71 clusters identified as having a tidal tail, we study the parameters of the 3 components solution. For the rest of the clusters, we adopt the parameters of the 2 components solution.

\citet{hu2021} used a similar approach and fitted a two components model using a Least Square Ellipse Fitting on 265 OCs from the membership catalogue by \citet{can18a}. A direct and systematic comparison of their results and ours is not appropriate because they investigate much smaller areas around each cluster compared to our study. However we checked some clusters individually. For instance, we find a tidal tail around NGC~752 (Fig \ref{f:tidal_tails}) with roughly the same orientation and eccentricity as them, according to their Fig. 1. We found an orientation of 6.146 degrees with the Galactic plane and an ellipticity of 0.776$\pm$0.002 while they found an angle of 4.697 degrees and an ellipticity of 0.615$\pm$0.342.

\subsection{Discussion}

\begin{figure}[h]
\centering
\includegraphics[width=\textwidth/2]{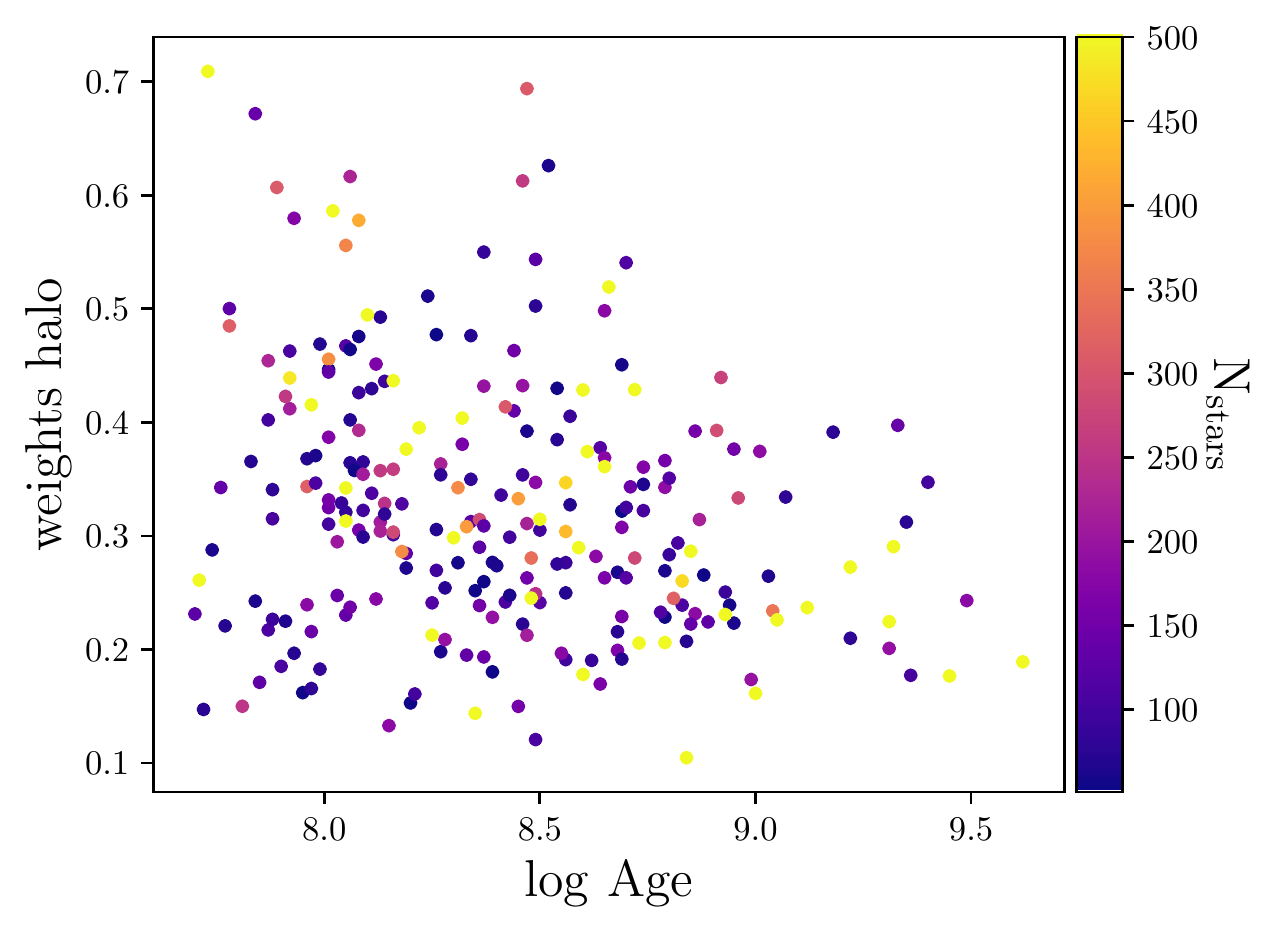}
\caption{Distribution of the weights of the second Gaussian component (associated to a halo structure) as a function of the age for the subsample of clusters which is best fitted with a two components model.}
\label{f:weights}
\end{figure}

The majority of the clusters have a significant corona. Among the clusters well described by a two components solution, there are 254 of them with a weight higher than 0.1 attributed to the corona. In Fig. \ref{f:weights} we show the weights of the halo as a function of the age for these clusters. We can see that even if young clusters can have halos with very various weights, we did not find any old clusters with a halo having a weight higher than 0.4. This indicates that as clusters grow old, less stars are part of the corona and that a higher proportion of their stars tend to concentrate in the cores of the clusters. This process might be connected to both the mass segregation and the cluster evaporation. Mass segregation tends to make the most massive stars of a cluster sink into its center and the less massive stars move towards its outskirts. On the other hand, if the cluster progressively evaporates, the outskirt stars are teared out from the corona as time goes by, thus reducing its weight as it is shown in Fig. \ref{f:weights}.

\citet{hu2021} and \citet{zhai2017} also fitted ellipses on the distribution of members projected on the plane of the sky. They noticed an increase of the ellipticity of the outer parts with age on a sample of 265 and 154 clusters respectively. Based on 31 OCs, \citet{chen2004} also noticed an increase of the circularity of the inner parts of OCs with age, especially at high altitudes. They attributed this process to internal dynamical relaxation process at stake in OCs, internal dynamics being able to shape clusters cores after $\sim$100 Myr while younger clusters inherit their shape from the cluster formation initial conditions. We considered the fitted parameters of the core and halo ellipses: length of the semi major axis, eccentricity and orientation. We looked for correlations with the Galactocentric radius, age and number of stars and we do not find any relevant trend. 

\begin{figure}[h]
\centering
\includegraphics[width=\textwidth/2]{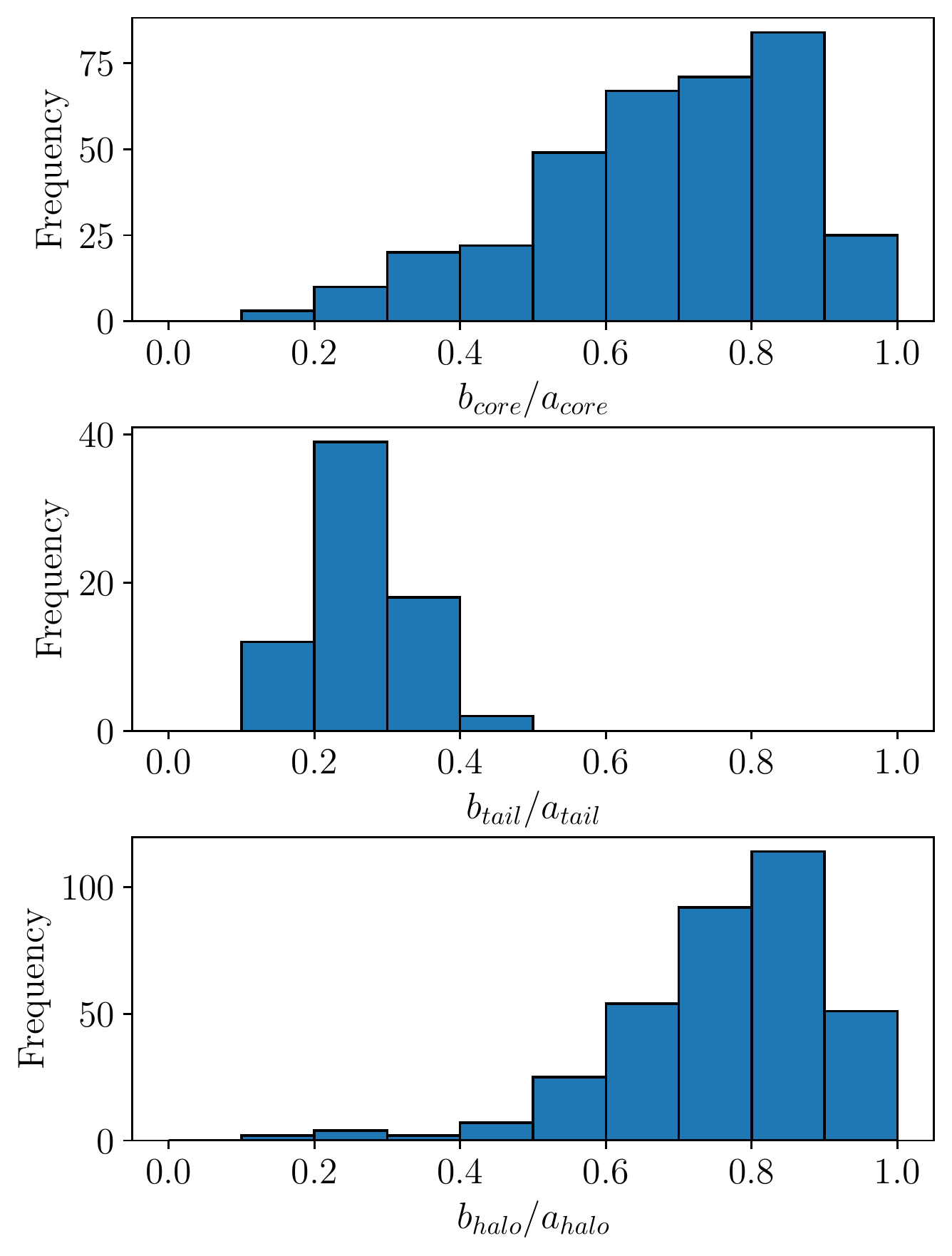}
\caption{Distribution of the axis ratio of each component.}
\label{f:histo_axis_ratio}
\end{figure}

As mentioned in Sect. \ref{sec:GMM_method}, we detected 71 tidal tails. We show on Fig. \ref{f:histo_axis_ratio} the distribution of the axis ratio of each fitted component for our whole sample of clusters. In the right hand panel, we show the distribution of the axis ratio of the cores. It shows that the vast majority of the clusters have a nearly circular core as the distribution peaks between 0.8 and 0.9. We checked the clusters with the most elliptical shape. They do not contain much stars in their center, making their ellipticity less reliable. In the middle panel is shown the distribution of the axis ratio of the tidal tails. They all are very elongated with $\sim$70\% of the identified tidal tails having an axis ratio lower than 0.3. Finally, the right hand panel shows the distribution of the axis ratio of the corona. In most of the cases, the corona is almost circular, and in the few cases where the ratio between the axis is lower than 0.5, again the number of stars in the halo is very low.

\begin{figure}[h]
\centering
\includegraphics[width=\textwidth/2]{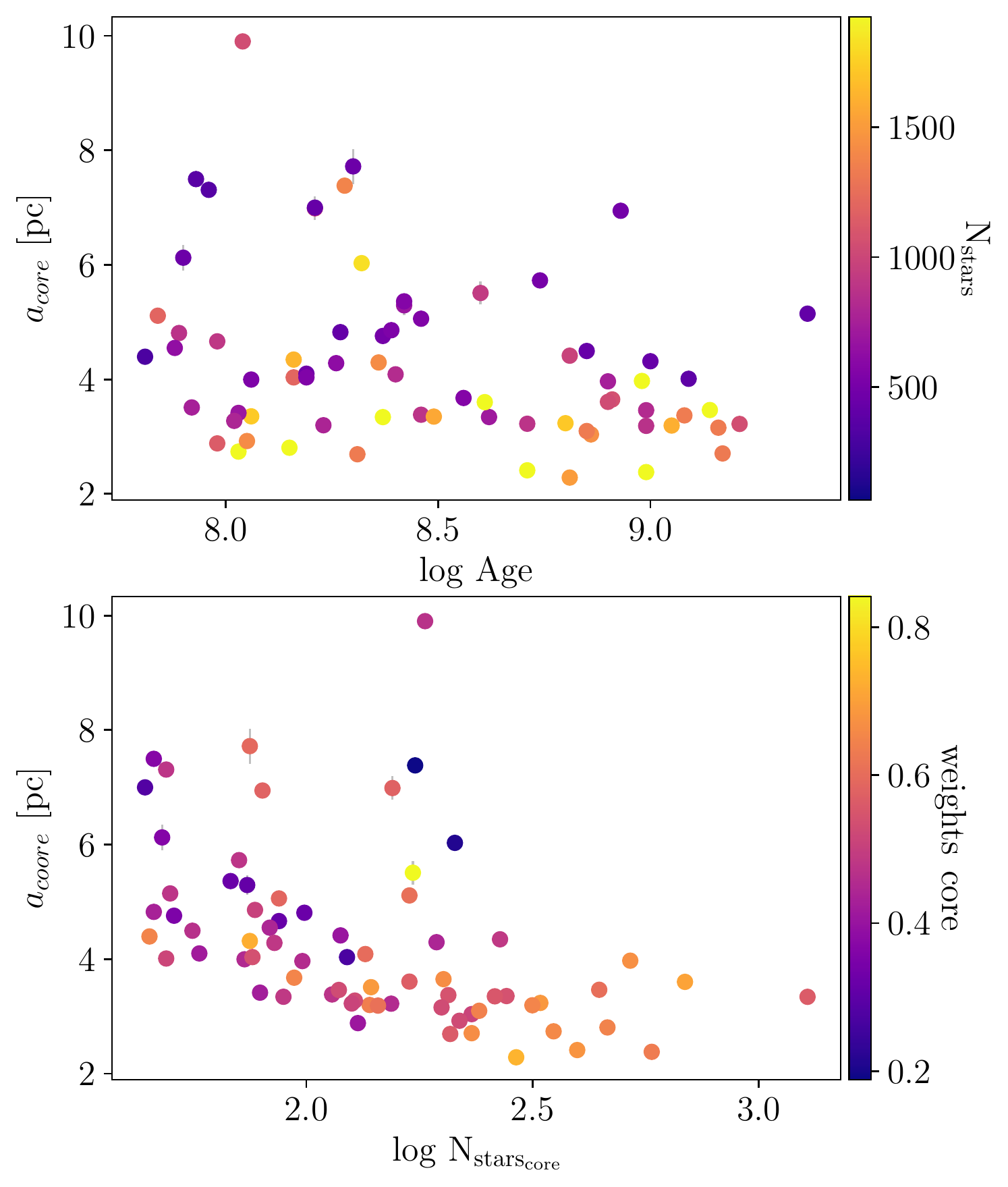}
\caption{Length of the semi major axis of the core of our clusters with age (top) and with the logarithm of the number of stars belonging to it (bottom) for the subsample of clusters with a tidal tail. On the top panel, the color bar stands for the total number of stars belonging to the cluster and on the bottom panel, it represents the relative weights of the core.}
\label{f:a_1_tidal}
\end{figure}

We can characterize the components of the 71 clusters in our sample showing a tidal tail by looking at the properties of their components. We show on Fig. \ref{f:a_1_tidal} the length of the core semi major axis as a function of clusters ages and of the logarithm of the number of stars belonging to it. 
The top panel shows that the clusters with tidal tail also follow the decreasing relation of the core radii with age found in Fig. \ref{fig:rt_rc_distribution} using all clusters. In the lower panel, it is shown that the length of the axis also has a decreasing dependence with the number of stars belonging to it. This means that even if populated clusters could be thought to have bigger cores, it seems on the contrary that their stronger gravitational binding make them much denser OCs. We looked for correlations between both tidal tails and halos semi-major axis lengths, eccentricities and orientations with the age of the clusters or with their locations in the Galaxy without finding any particular trend. 

Sixteen tidal tails have already been characterised in the literature. Eight of them are part of our sample : Coma~Berenices, Ruprecht~147, Praesepe, Blanco~1, NGC~752, NGC~7092, NGC~2516 and Platais~9. Our study identifies tidal tails for Coma~Berenices, Blanco~1, NGC~752, NGC~7092 and NGC~2516 previously found respectively by \citet{tang2019}, \citet{zhang2020}, \cite{bha2021} and \citet{mei21}. We found the same orientations as these authors. The tidal tails of Ruprecht~147, Praesepe, and Platais~9 were characterised respectively by \citet{yeh2019}, \citet{roe_praesepe}, \citet{gao2020a} and \citet{mei21} who found them to be roughly aligned with the line of sight. We cannot detect such tidal tails owing to our 2D analysis of the distribution of stars projected on the sky.

\citet{mei21} identified extended stellar populations similar to tidal structure in nine out of 10 OCs. They attributed these structures to the imprint of the parent molecular cloud relic structure for the clusters younger than 50 Myr and to stripped cluster stars for the clusters with age $\gtrsim$ 100 Myr. \citet{pang2021} also found elongated shapes in 8 of their 13 OCs sample in the form of filamentary like substructures, reminiscent of the star formation history of the cluster for the clusters younger than 50 Myr and to tidal stripping for the oldest ones (NGC~2516, Blanco~1, Coma~Berenices, NGC~6633 and Ruprecht~147). We find the same structures for the clusters we have in common (with the exception of Ruprecht~147). According to \citet{lada2003}, \citet{bon1998} and \citet{bas2009}, in only a few million years, a cluster would reach a state of equilibrium and get rid of its star distribution inherited from the star formation process. As we do not have any clusters younger than 50 Myr in our sample, the vast majority of the tidal tails in our sample can be attributed to dynamical effects and to tidal stripping.

\section{Mass segregation}\label{sec:mass_segregation}
According to the standard view, mass segregation in OCs is believed to increase with age \citep{kro1995,dib2018}. Old clusters are therefore believed to be more frequently mass segregated than young clusters. Thanks to our membership analysis over extended regions of OCs of various age, this is an hypothesis that we can try to verify.

\subsection{Method}

In order to measure the degree of mass segregation we applied the method proposed by \citet{all2009} and widely used to quantify and detect mass segregation in stellar clusters \citep{nony2021, dib2018, plu2018, zun2019}.  This method works by comparing the length of the minimum spanning tree (MST) of the most massive stars of a cluster with the length of the MST of a set of the same number of randomly chosen stars. A MST of a set of points is the path connecting all the points, with the shortest pathlength possible and without any closed loops. In a given set of points, only one MST can be drawn. We computed the MST by using the \texttt{csgraph} routine implemented in the scipy python module \citep{scipy}. In all the cases, we drew the minimum spanning tree in the same set of coordinates as defined in Eq. \ref{eq:projected_coordinated}. The mass segregation ratio (MSR) $\Lambda_{MSR}$ is then defined as follows : 

\begin{align}
\begin{aligned}
\Lambda_{MSR} (N) = \frac{<l_{random}>}{l_{massive}} \pm \frac{\sigma_{random}}{l_{massive}},
\end{aligned}\label{eq:MSR}
\end{align}
with $<l_{random}>$ being the average length of the MST of $N$ randomly chosen stars and $l_{massive}$ the length of the MST of the $N$ most massive stars. The average length $<l_{random}>$ was calculated over 100 iterations where at each iteration we draw a different subsample of random stars allowing us to calculate at the same time $\sigma_{random}$, the standard deviation of the length of the MST of these $N$ stars. We used the $G$ magnitude of the stars as a proxy for the mass. The mass segregation ratio $\Lambda_{MSR}$ was always calculated for a subset of $N$ stars. A result of $\Lambda_{MSR}$ greater than 1 means that the $N$ most massive stars are more concentrated compared to a random sample and therefore that the cluster shows a sign of mass segregation. We conducted this analysis for all the clusters containing more than 50 stars. In each cluster we calculated the MSR starting at $N=5$ up to the number of cluster members. We started at $N=5$ because for lower N the value is not statistically significant. For clusters with more than 100 stars, we stopped at $N=100$ since the MSR only shows a gradual decrease to reach unity.

\begin{figure}[h]
\centering
\includegraphics[width=\textwidth/2]{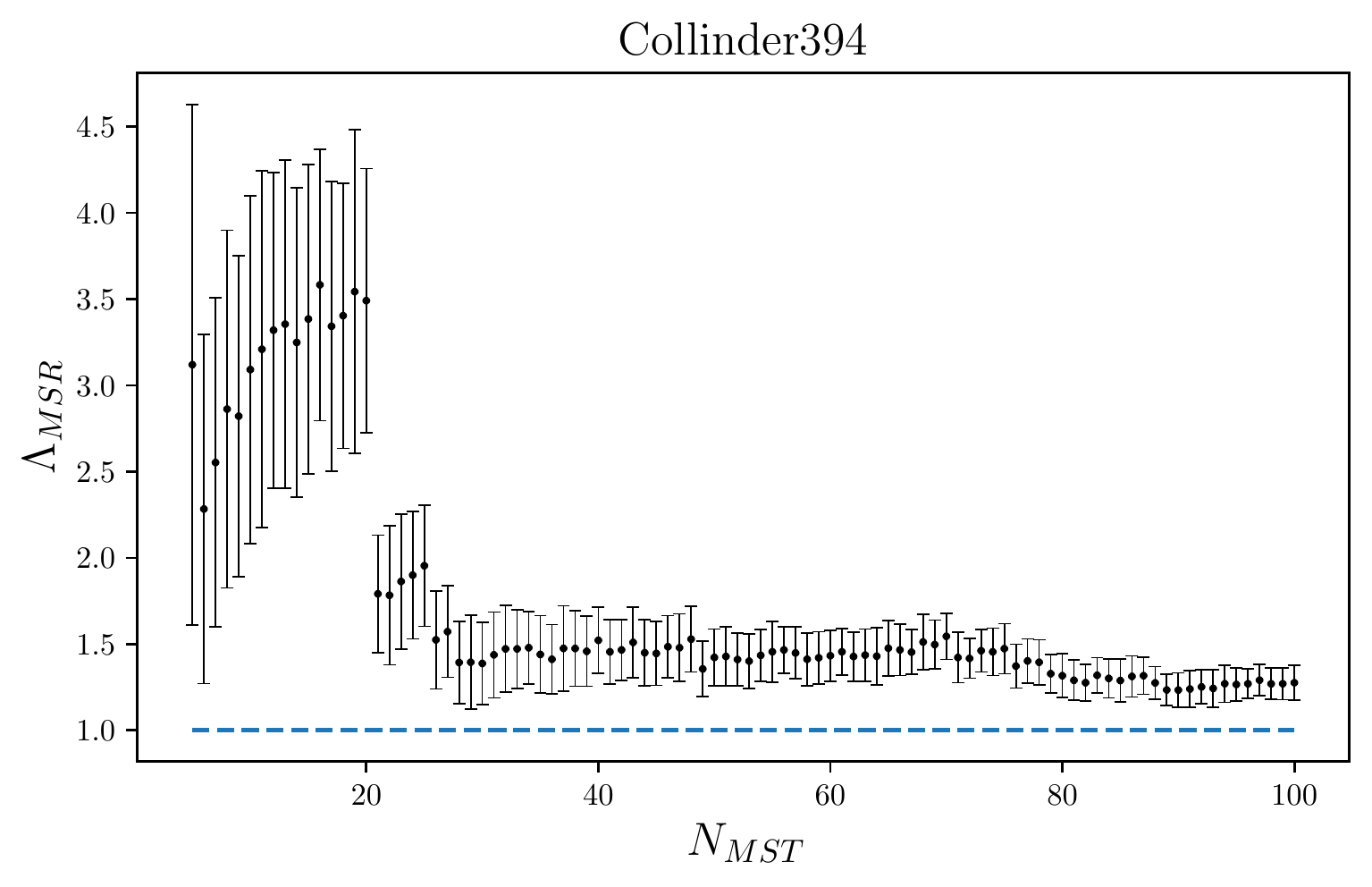}
\caption{Mass segregation ratio $\Lambda_{MSR}(N)$ for the cluster Collinder~394 as a function of the number of stars used to draw the MST. The blue dotted line shows the limit $\Lambda_{MSR}=1$ after which stars do not show any signs of mass segregation.}
\label{f:MS_Collinder_394}
\end{figure}

Figure \ref{f:MS_Collinder_394} shows the mass segregation ratio for the cluster Collinder~394 for increasing values of $N$. There are several degrees of mass segregation. First the 20 most massive stars have a value for $\Lambda_{MSR} \sim$3.2. There is then a drop of $\Lambda_{MSR}$ with a plateau for $20<N<26$ around a value of $\sim$1.7. The mass segregation ratio then drops to 1.4 and progressively decreases. This analysis tells us that in Collinder 394, the 20 most massive stars are 3.2 times closer to each other compared to the typical separation of 20 random stars in the cluster and that the 25 most massive stars of cluster are 70\% more concentrated compared to any set of 25 members. After that, the rest of the stars progressively approach $\Lambda_{MSR} \sim$1.

\subsection{Discussion}

\begin{figure}[h]
\centering
\includegraphics[width=\textwidth/2]{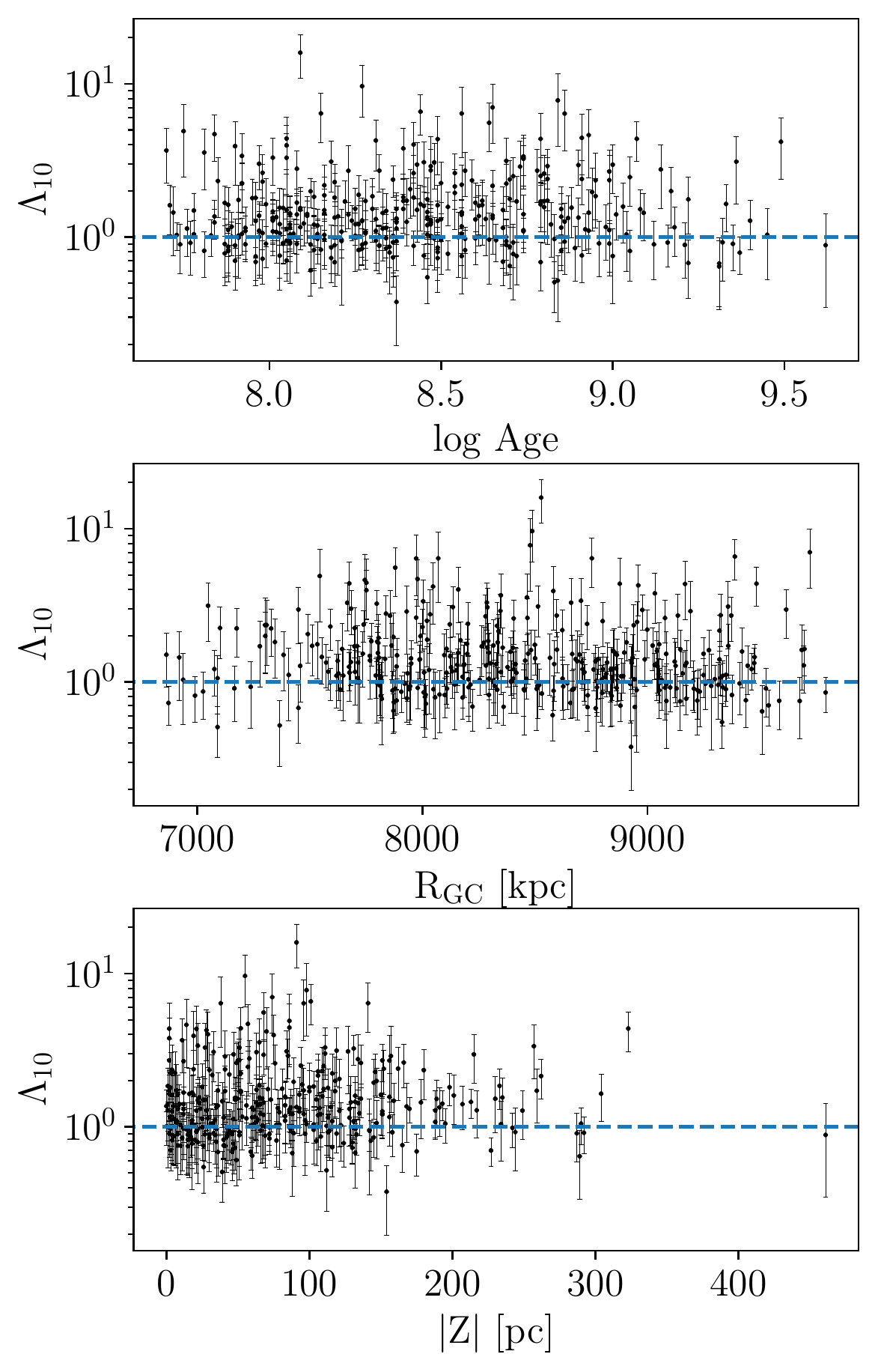}
\caption{Mass segregation ratio $\Lambda_{10}$ of the 10 most massive stars of each cluster as a function of OCs ages (top), Galactocentric radii (middle) and absolute value of the altitude above the galactic mid-plane (bottom). The blue dotted line shows the limit $\Lambda_{MSR}=1$.}
\label{f:age_lambda}
\end{figure}

\citet{mas2010} studied the very early stages of clusters through N-body simulations and noted that the 10 most massive stars of a cluster quickly formed a very concentrated system once the clusters are formed. We represent in Fig. \ref{f:age_lambda} the distribution of the mass segregation ratio of the 10th most massive stars ($\Lambda_{10}$) as a function of their parent cluster ages, Galactocentric radii and altitude above (or below) the galactic plane. We do not notice any particular trend regarding the evolution of mass segregation with these parameters even though we would expect an increase of the mass segregation with age \citep{dib2018}. For consistency, we also checked if some trend appeared when looking at the MSR of the fifth to the twentieth most massive stars but that was not conclusive. 

The lack of a net relation between $\Lambda_{10}$ with age might be explained by the fact that OCs formed with very different levels of mass segregation \citep{dib2018}. For instance, the very young cluster Trapezium, the core of the Orion Nebula Cluster, shows evidence of mass segregation even though its age is $\sim$1 Myr \citep{bon1998, all2009}. This was referred to as primordial mass segregation \citep{degrijs2003}. Mass segregation in young clusters was first thought to be caused by the initial conditions of the cluster formation, but recent N-body simulations suggested that mass segregation occurs on time scales of the order of a few Myr. This implies that clusters younger than their dynamical relaxation time can show signs of mass segregation. This could be due to dynamical interactions through the merging of smaller substructures. In this scenario, clusters are born with a significant amount of clumps. Each of these small clumps can then mass segregate in short time scales through dynamical interactions. The merging of these multiple clumps later gives birth to a cluster that inherited the substructure's segregation \citep{mcmi2007, all2009b, all2010, mas2010}. This gives a more complex view of what is expected to be observed in our sample of OCs.

\begin{figure}[h]
\centering
\includegraphics[width=\textwidth/2]{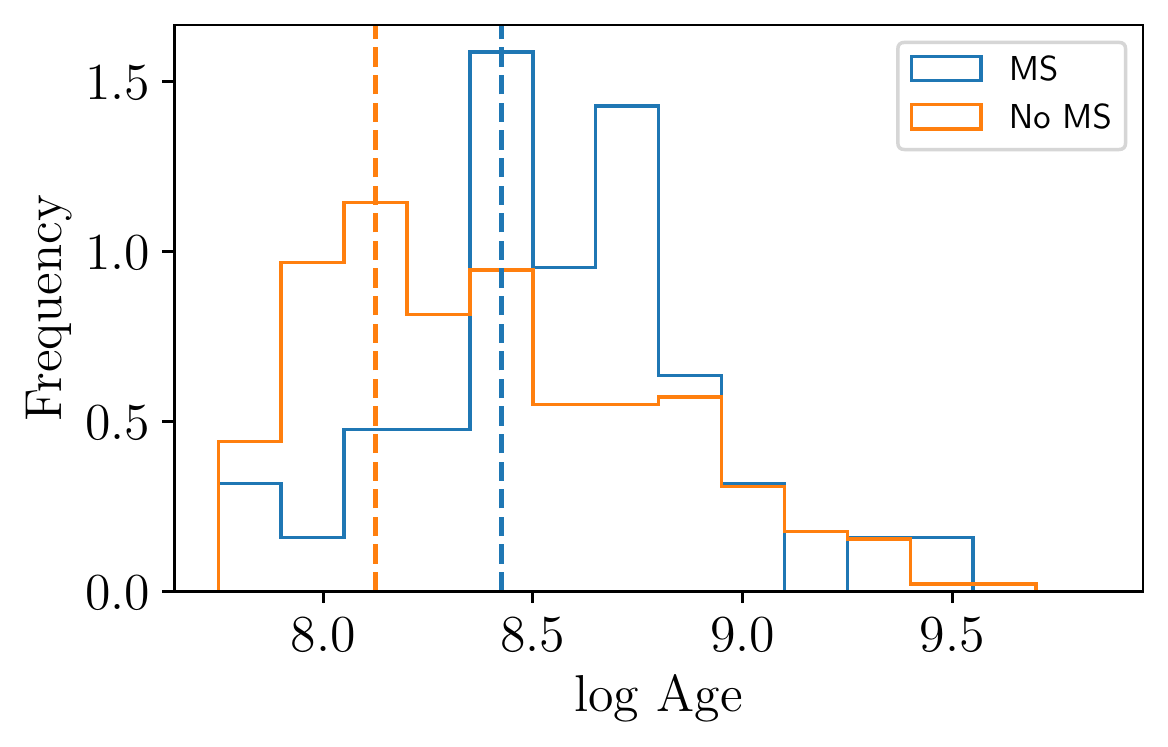}
\caption{Age distribution of two subsamples of OCs : one where more than 10\% of the stars have $\Lambda_{MSR}>2$ (in blue) and one where less than 10\% of the stars have $\Lambda_{MSR}>2$ (in orange). The vertical dotted line shows the mode of each distribution. }
\label{f:ms_age_distrib}
\end{figure}

In order to investigate in a different way the age dependency of MS, we measured the proportion of stars per cluster having $\Lambda_{MSR}>2$. We defined two subsamples : one where more than 10\% of the stars have $\Lambda_{MSR}>2$ and the other one being the complementary subsample. For example, we can see in Fig. \ref{f:MS_Collinder_394} that in Collinder~394, 20 stars have a MSR higher than 2. As it counts 703 members, only 2\% of Collinder~394 stars have $\Lambda_{MSR}>2$ and Collinder~394 is therefore part of the complementary subsample where less than 10\% of the clusters members are highly mass segregated. The age distribution of these two subsamples is shown in Fig \ref{f:ms_age_distrib}. Even if no trend between $\Lambda_{10}$ and the cluster ages was noticeable in Fig \ref{f:age_lambda}, it is clear here that OCs which have a large proportion of stars showing a strong mass segregation are -in average- older than the clusters with few stars highly mass segregated. 

This could be related with the signs of evaporation observed in Sect \ref{sec:GMM}. In Fig. \ref{f:weights}, we noted that old clusters have -in average- less stars in their halo than young ones. As old clusters are proportionally more mass segregated than young ones, more small stars should be pushed to the outskirts of the clusters, increasing the relative weights of the halo. As we observe the opposite, it could indicate that cluster evaporation process is very efficient compared to mass segregation.

\section{Conclusion}\label{sec:conclusion}
We developed a methodology able to identify members in the peripheral area of OCs, up to 50 pc from their center. The method is based on the unsupervised clustering algorithm HDBSCAN which can detect overdensities in the astrometric space ($\mu_{\alpha^{*}}, \mu_{\delta}, \varpi$) even in datasets of varying density. We applied this method on 467 OCs from CAN+20 closer than 1.5 kpc and older than 50 Myr. We report memberships for 389 OCs, the 78 remaining ones being too embedded in their field for our method to properly disentangle them from their neighbors. For the vast majority of clusters, we identify many more members than previously known. For COIN-Gaia 13, a very extended cluster, we tried to increase our search radius and recovered members up to 150 pc from the clusters centers. This highlights that studies focused on small samples of clusters should look for members even at large distances from the centers of the clusters, like \cite{mei21} or \cite{car19}.

We identify vast coronae around almost all the clusters, reaching in most cases the maximum radius of the investigated area. We also identify tidal tails for $\sim$71 OCs. Previous detections of corona or tidal tails of OCs were focused on smaller samples of nearby clusters. Since we worked with the 2D projected distribution of stars, we were able to perform a systematic study of OCs at large heliocentric distances, and to multiply by more than 4 the number of identified clusters with a tidal tail. 

The primary goal of this paper is to determine the structural parameters of the clusters for which we obtained new members at large radius. To do so, we fitted the radial density profile of each cluster with more than 100 members with a King function in order to study their core and tidal radii. We find similar core radii to the ones published in previous studies but as we find members at much larger radius than in previous studies, we also find much larger tidal radii. The distribution of the fitted core radii shows a concentration between 1 and 2.5 pc regardless of the age or of the number of cluster members. We see that older clusters tend to have smaller $R_c$ compared to young clusters with values converging towards 1.85 pc at the age of 1 Gyr. The tidal radii peaks around 30 pc, but more importantly seems to increase with age in what could be a sign of mass segregation, dissolution or a combination of both processes. The tidal radius distribution could be biased due to the limit of 50 pc that we used to query the \emph{Gaia} EDR3 catalog. A fraction of the investigated clusters may extend to larger radii than this, as shown in Sect. \ref{sec:new_memberships} with the example of COIN-Gaia 13.

We fitted GMMs on the spatial distribution of members projected on a plane perpendicular to the celestial sphere. This is particularly suited for the clusters for which we detected an elongated tidal tail as the King function previously used assumes a spherical distribution of members. We used a three components GMM on the 71 clusters with a tidal tail, one component representing the core of the cluster, one the tidal tail and one the corona. For the other clusters, we adopted a two component GMM with a core and a halo. We looked for correlations between the parameters of the fitted Gaussians with the characteristics of the clusters (i.e. their age, location, number of members, etc). For the 71 clusters of our sample with a tidal tail, we observe that old ones are more prone to have small cores than young ones. We also note that the relative weight of the corona of old clusters was in average lower than for young ones. This implies that with an increasing age, the proportion of stars in the cluster halos decreases, either because stars move to the center of the cluster or because outer stars are ejected from the cluster. 

We applied the method proposed by \cite{all2009} to measure the degree of mass segregation of our sample of OCs. We do not notice any trend of the mass segregation ratio, measured through the 10 most massive stars, with age, Galactocentric distance or with the altitude of the cluster above the Galactic mid-plane. However, clusters having a significant number of stars with a strong mass segregation ratio, are on average older than clusters with few stars strongly mass segregated. Coupled with a lower proportion of stars populating the clusters halos, this highlights the fact that the various physical processes in play in the disruption of clusters acts on shorter time scales than mass segregation. 

\begin{acknowledgements}
This work has made use of data from the European Space Agency (ESA) mission \emph{Gaia} (\url{http://www.cosmos.esa.int/Gaia}), processed by the \emph{Gaia} Data Processing and Analysis Consortium (DPAC, \url{http://www.cosmos.esa.int/web/Gaia/dpac/consortium}). We acknowledge the \emph{Gaia} Project Scientist Support Team and the \emph{Gaia} DPAC. Funding for the DPAC has been provided by national institutions, in particular the institutions participating in the \emph{Gaia} Multilateral Agreement.
This research made extensive use of the SIMBAD database, and the VizieR catalogue access tool, operated at the CDS, Strasbourg, France, and of NASA Astrophysics Data System Bibliographic Services.
This research has made use of Astropy \citep{Astropy2013}, Topcat \citep{Taylor2005}.

Y.T., C.S., and L.C. acknowledge support from "programme national de physique stellaire" (PNPS) and from the "programme national cosmologie et galaxies" (PNCG) of CNRS/INSU. L.C. acknowledges the support of the postdoc fellowship from French Centre National d’Etudes Spatiales (CNES). This work was (partially) supported by the Spanish Ministry of Science, Innovation and University (MICIU/FEDER, UE) through grant RTI2018-095076-B-C21, and the Institute of Cosmos Sciences University of Barcelona (ICCUB, Unidad de Excelencia ’Mar\'{\i}a de Maeztu’) through grant CEX2019-000918-M. J.O. acknowledges financial support from: i)  from the European Research Council (ERC) under the European Union’s Horizon 2020 research and innovation program (grant agreement No 682903, P.I. H. Bouy), ii) the French State in the framework of the ”Investments for the future” Program, IdEx Bordeaux, reference ANR-10-IDEX-03-02, and iii) the Agencia Estatal de Investigaci\'on of the Ministerio de Ciencia, Innovaci\'on y Universidades  through project PID2019-109522GB-C53.
\end{acknowledgements}


\bibliographystyle{aa} 
\bibliography{biblio}

\begin{thebibliography}{81}
\expandafter\ifx\csname natexlab\endcsname\relax\def\natexlab#1{#1}\fi

\bibitem[{{Allison} {et~al.}(2009{\natexlab{a}}){Allison}, {Goodwin}, {Parker},
  {de Grijs}, {Portegies Zwart}, \& {Kouwenhoven}}]{all2009b}
{Allison}, R.~J., {Goodwin}, S.~P., {Parker}, R.~J., {et~al.}
  2009{\natexlab{a}}, \apjl, 700, L99

\bibitem[{{Allison} {et~al.}(2010){Allison}, {Goodwin}, {Parker}, {Portegies
  Zwart}, \& {de Grijs}}]{all2010}
{Allison}, R.~J., {Goodwin}, S.~P., {Parker}, R.~J., {Portegies Zwart}, S.~F.,
  \& {de Grijs}, R. 2010, \mnras, 407, 1098

\bibitem[{{Allison} {et~al.}(2009{\natexlab{b}}){Allison}, {Goodwin}, {Parker},
  {Portegies Zwart}, {de Grijs}, \& {Kouwenhoven}}]{all2009}
{Allison}, R.~J., {Goodwin}, S.~P., {Parker}, R.~J., {et~al.}
  2009{\natexlab{b}}, \mnras, 395, 1449

\bibitem[{{Alves} {et~al.}(2020){Alves}, {Zucker}, {Goodman}, {Speagle},
  {Meingast}, {Robitaille}, {Finkbeiner}, {Schlafly}, \& {Green}}]{alves2020}
{Alves}, J., {Zucker}, C., {Goodman}, A.~A., {et~al.} 2020, \nat, 578, 237

\bibitem[{{Angelo} {et~al.}(2021){Angelo}, {Corradi}, {Santos}, {Maia}, \&
  {Ferreira}}]{ang21}
{Angelo}, M.~S., {Corradi}, W.~J.~B., {Santos}, J.~F.~C., J., {Maia}, F.~F.~S.,
  \& {Ferreira}, F.~A. 2021, \mnras, 500, 4338

\bibitem[{{Artyukhina} \& {Kholopov}(1964)}]{art1964}
{Artyukhina}, N.~M. \& {Kholopov}, P.~N. 1964, \sovast, 7, 840

\bibitem[{{Astropy Collaboration} {et~al.}(2013){Astropy Collaboration},
  {Robitaille}, {Tollerud}, {Greenfield}, {Droettboom}, {Bray}, {Aldcroft},
  {Davis}, {Ginsburg}, {Price-Whelan}, {Kerzendorf}, {Conley}, {Crighton},
  {Barbary}, {Muna}, {Ferguson}, {Grollier}, {Parikh}, {Nair}, {Unther},
  {Deil}, {Woillez}, {Conseil}, {Kramer}, {Turner}, {Singer}, {Fox}, {Weaver},
  {Zabalza}, {Edwards}, {Azalee Bostroem}, {Burke}, {Casey}, {Crawford},
  {Dencheva}, {Ely}, {Jenness}, {Labrie}, {Lim}, {Pierfederici}, {Pontzen},
  {Ptak}, {Refsdal}, {Servillat}, \& {Streicher}}]{Astropy2013}
{Astropy Collaboration}, {Robitaille}, T.~P., {Tollerud}, E.~J., {et~al.} 2013,
  \aap, 558, A33

\bibitem[{{Bailer-Jones}(2015)}]{bai2015}
{Bailer-Jones}, C. A.~L. 2015, \pasp, 127, 994

\bibitem[{{Bastian} {et~al.}(2009){Bastian}, {Gieles}, {Ercolano}, \&
  {Gutermuth}}]{bas2009}
{Bastian}, N., {Gieles}, M., {Ercolano}, B., \& {Gutermuth}, R. 2009, \mnras,
  392, 868

\bibitem[{{Baumgardt} \& {Kroupa}(2007)}]{baumgardt2007}
{Baumgardt}, H. \& {Kroupa}, P. 2007, \mnras, 380, 1589

\bibitem[{{Bhattacharya} {et~al.}(2021){Bhattacharya}, {Agarwal}, {Rao}, \&
  {Vaidya}}]{bha2021}
{Bhattacharya}, S., {Agarwal}, M., {Rao}, K.~K., \& {Vaidya}, K. 2021, \mnras
  [\eprint[arXiv]{2105.06108}]

\bibitem[{{Binney} \& {Tremaine}(1987)}]{binney_tremaine}
{Binney}, J. \& {Tremaine}, S. 1987, {Galactic dynamics}

\bibitem[{{Bonnell} \& {Davies}(1998)}]{bon1998}
{Bonnell}, I.~A. \& {Davies}, M.~B. 1998, \mnras, 295, 691

\bibitem[{Campello {et~al.}(2013)Campello, Moulavi, \& Sander}]{HDBSCAN}
Campello, R. J. G.~B., Moulavi, D., \& Sander, J. 2013, in Advances in
  Knowledge Discovery and Data Mining, ed. J.~Pei, V.~S. Tseng, L.~Cao,
  H.~Motoda, \& G.~Xu (Berlin, Heidelberg: Springer Berlin Heidelberg),
  160--172

\bibitem[{{Cantat-Gaudin} {et~al.}(2020){Cantat-Gaudin}, {Anders},
  {Castro-Ginard}, {Jordi}, {Romero-G{\'o}mez}, {Soubiran}, {Casamiquela},
  {Tarricq}, {Moitinho}, {Vallenari}, {Bragaglia}, {Krone-Martins}, \&
  {Kounkel}}]{can20b}
{Cantat-Gaudin}, T., {Anders}, F., {Castro-Ginard}, A., {et~al.} 2020, \aap,
  640, A1

\bibitem[{{Cantat-Gaudin} {et~al.}(2018{\natexlab{a}}){Cantat-Gaudin}, {Jordi},
  {Vallenari}, {Bragaglia}, {Balaguer-N{\'u}{\~n}ez}, {Soubiran}, {Bossini},
  {Moitinho}, {Castro-Ginard}, {Krone-Martins}, {Casamiquela}, {Sordo}, \&
  {Carrera}}]{can18}
{Cantat-Gaudin}, T., {Jordi}, C., {Vallenari}, A., {et~al.} 2018{\natexlab{a}},
  \aap, 618, A93

\bibitem[{{Cantat-Gaudin} {et~al.}(2018{\natexlab{b}}){Cantat-Gaudin},
  {Vallenari}, {Sordo}, {Pensabene}, {Krone-Martins}, {Moitinho}, {Jordi},
  {Casamiquela}, {Balaguer-N{\'u}nez}, {Soubiran}, \& {Brouillet}}]{can18a}
{Cantat-Gaudin}, T., {Vallenari}, A., {Sordo}, R., {et~al.} 2018{\natexlab{b}},
  \aap, 615, A49

\bibitem[{{Carrera} {et~al.}(2019){Carrera}, {Pasquato}, {Vallenari},
  {Balaguer-N{\'u}{\~n}ez}, {Cantat-Gaudin}, {Mapelli}, {Bragaglia}, {Bossini},
  {Jordi}, {Galad{\'\i}-Enr{\'\i}quez}, \& {Solano}}]{car19}
{Carrera}, R., {Pasquato}, M., {Vallenari}, A., {et~al.} 2019, \aap, 627, A119

\bibitem[{{Castro-Ginard} {et~al.}(2020){Castro-Ginard}, {Jordi}, {Luri},
  {{\'A}lvarez Cid-Fuentes}, {Casamiquela}, {Anders}, {Cantat-Gaudin},
  {Mongui{\'o}}, {Balaguer-N{\'u}{\~n}ez}, {Sol{\`a}}, \& {Badia}}]{cas20}
{Castro-Ginard}, A., {Jordi}, C., {Luri}, X., {et~al.} 2020, \aap, 635, A45

\bibitem[{{Castro-Ginard} {et~al.}(2019){Castro-Ginard}, {Jordi}, {Luri},
  {Cantat-Gaudin}, \& {Balaguer-N{\'u}{\~n}ez}}]{cas19}
{Castro-Ginard}, A., {Jordi}, C., {Luri}, X., {Cantat-Gaudin}, T., \&
  {Balaguer-N{\'u}{\~n}ez}, L. 2019, \aap, 627, A35

\bibitem[{{Castro-Ginard} {et~al.}(2018){Castro-Ginard}, {Jordi}, {Luri},
  {Julbe}, {Morvan}, {Balaguer-N{\'u}{\~n}ez}, \& {Cantat-Gaudin}}]{cas18}
{Castro-Ginard}, A., {Jordi}, C., {Luri}, X., {et~al.} 2018, \aap, 618, A59

\bibitem[{{Chen} {et~al.}(2004){Chen}, {Chen}, \& {Shu}}]{chen2004}
{Chen}, W.~P., {Chen}, C.~W., \& {Shu}, C.~G. 2004, \aj, 128, 2306

\bibitem[{{de Grijs} {et~al.}(2003){de Grijs}, {Gilmore}, \&
  {Johnson}}]{degrijs2003}
{de Grijs}, R., {Gilmore}, G.~F., \& {Johnson}, R. 2003, in Astronomical
  Society of the Pacific Conference Series, Vol. 296, New Horizons in Globular
  Cluster Astronomy, ed. G.~{Piotto}, G.~{Meylan}, S.~G. {Djorgovski}, \&
  M.~{Riello}, 207

\bibitem[{{de La Fuente Marcos}(1996)}]{delafuente1996}
{de La Fuente Marcos}, R. 1996, \aap, 314, 453

\bibitem[{{Dias} {et~al.}(2002){Dias}, {Alessi}, {Moitinho}, \&
  {L{\'e}pine}}]{dias2002}
{Dias}, W.~S., {Alessi}, B.~S., {Moitinho}, A., \& {L{\'e}pine}, J.~R.~D. 2002,
  \aap, 389, 871

\bibitem[{{Dib} {et~al.}(2018){Dib}, {Schmeja}, \& {Parker}}]{dib2018}
{Dib}, S., {Schmeja}, S., \& {Parker}, R.~J. 2018, \mnras, 473, 849

\bibitem[{{Dinnbier} \& {Kroupa}(2020{\natexlab{a}})}]{din2020}
{Dinnbier}, F. \& {Kroupa}, P. 2020{\natexlab{a}}, \aap, 640, A84

\bibitem[{{Dinnbier} \& {Kroupa}(2020{\natexlab{b}})}]{dinn2020b}
{Dinnbier}, F. \& {Kroupa}, P. 2020{\natexlab{b}}, \aap, 640, A85

\bibitem[{Ester {et~al.}(1996)Ester, Kriegel, Sander, \& Xu}]{DBSCAN}
Ester, M., Kriegel, H.-P., Sander, J., \& Xu, X. 1996, in Proceedings of the
  Second International Conference on Knowledge Discovery and Data Mining,
  KDD'96 (AAAI Press), 226–231

\bibitem[{{Fabricius} {et~al.}(2021){Fabricius}, {Luri}, {Arenou}, {Babusiaux},
  {Helmi}, {Muraveva}, {Reyl{\'e}}, {Spoto}, {Vallenari}, {Antoja}, {Balbinot},
  {Barache}, {Bauchet}, {Bragaglia}, {Busonero}, {Cantat-Gaudin}, {Carrasco},
  {Diakit{\'e}}, {Fabrizio}, {Figueras}, {Garcia-Gutierrez}, {Garofalo},
  {Jordi}, {Kervella}, {Khanna}, {Leclerc}, {Licata}, {Lambert}, {Marrese},
  {Masip}, {Ramos}, {Robichon}, {Robin}, {Romero-G{\'o}mez}, {Rubele}, \&
  {Weiler}}]{edr3_catalogue}
{Fabricius}, C., {Luri}, X., {Arenou}, F., {et~al.} 2021, \aap, 649, A5

\bibitem[{{Foreman-Mackey} {et~al.}(2013){Foreman-Mackey}, {Hogg}, {Lang}, \&
  {Goodman}}]{emcee}
{Foreman-Mackey}, D., {Hogg}, D.~W., {Lang}, D., \& {Goodman}, J. 2013, \pasp,
  125, 306

\bibitem[{{Gaia Collaboration} {et~al.}(2018){Gaia Collaboration}, {Brown},
  {Vallenari}, {Prusti}, {de Bruijne}, {Babusiaux}, {Bailer-Jones}, {Biermann},
  {Evans}, {Eyer}, {Jansen}, {Jordi}, {Klioner}, {Lammers}, {Lindegren},
  {Luri}, {Mignard}, {Panem}, {Pourbaix}, {Randich}, {Sartoretti}, {Siddiqui},
  {Soubiran}, {van Leeuwen}, {Walton}, {Arenou}, {Bastian}, {Cropper},
  {Drimmel}, {Katz}, {Lattanzi}, {Bakker}, {Cacciari}, {Casta{\~n}eda},
  {Chaoul}, {Cheek}, {De Angeli}, {Fabricius}, {Guerra}, {Holl}, {Masana},
  {Messineo}, {Mowlavi}, {Nienartowicz}, {Panuzzo}, {Portell}, {Riello},
  {Seabroke}, {Tanga}, {Th{\'e}venin}, {Gracia-Abril}, {Comoretto},
  {Garcia-Reinaldos}, {Teyssier}, {Altmann}, {Andrae}, {Audard},
  {Bellas-Velidis}, {Benson}, {Berthier}, {Blomme}, {Burgess}, {Busso},
  {Carry}, {Cellino}, {Clementini}, {Clotet}, {Creevey}, {Davidson}, {De
  Ridder}, {Delchambre}, {Dell'Oro}, {Ducourant},
  {Fern{\'a}ndez-Hern{\'a}ndez}, {Fouesneau}, {Fr{\'e}mat}, {Galluccio},
  {Garc{\'\i}a-Torres}, {Gonz{\'a}lez-N{\'u}{\~n}ez}, {Gonz{\'a}lez-Vidal},
  {Gosset}, {Guy}, {Halbwachs}, {Hambly}, {Harrison}, {Hern{\'a}ndez},
  {Hestroffer}, {Hodgkin}, {Hutton}, {Jasniewicz}, {Jean-Antoine-Piccolo},
  {Jordan}, {Korn}, {Krone-Martins}, {Lanzafame}, {Lebzelter}, {L{\"o}ffler},
  {Manteiga}, {Marrese}, {Mart{\'\i}n-Fleitas}, {Moitinho}, {Mora}, {Muinonen},
  {Osinde}, {Pancino}, {Pauwels}, {Petit}, {Recio-Blanco}, {Richards},
  {Rimoldini}, {Robin}, {Sarro}, {Siopis}, {Smith}, {Sozzetti}, {S{\"u}veges},
  {Torra}, {van Reeven}, {Abbas}, {Abreu Aramburu}, {Accart}, {Aerts},
  {Altavilla}, {{\'A}lvarez}, {Alvarez}, {Alves}, {Anderson}, {Andrei},
  {Anglada Varela}, {Antiche}, {Antoja}, {Arcay}, {Astraatmadja}, {Bach},
  {Baker}, {Balaguer-N{\'u}{\~n}ez}, {Balm}, {Barache}, {Barata}, {Barbato},
  {Barblan}, {Barklem}, {Barrado}, {Barros}, {Barstow}, {Bartholom{\'e}
  Mu{\~n}oz}, {Bassilana}, {Becciani}, {Bellazzini}, {Berihuete}, {Bertone},
  {Bianchi}, {Bienaym{\'e}}, {Blanco-Cuaresma}, {Boch}, {Boeche}, {Bombrun},
  {Borrachero}, {Bossini}, {Bouquillon}, {Bourda}, {Bragaglia}, {Bramante},
  {Breddels}, {Bressan}, {Brouillet}, {Br{\"u}semeister}, {Brugaletta},
  {Bucciarelli}, {Burlacu}, {Busonero}, {Butkevich}, {Buzzi}, {Caffau},
  {Cancelliere}, {Cannizzaro}, {Cantat-Gaudin}, {Carballo}, {Carlucci},
  {Carrasco}, {Casamiquela}, {Castellani}, {Castro-Ginard}, {Charlot},
  {Chemin}, {Chiavassa}, {Cocozza}, {Costigan}, {Cowell}, {Crifo}, {Crosta},
  {Crowley}, {Cuypers}, {Dafonte}, {Damerdji}, {Dapergolas}, {David}, {David},
  {de Laverny}, {De Luise}, {De March}, {de Martino}, {de Souza}, {de Torres},
  {Debosscher}, {del Pozo}, {Delbo}, {Delgado}, {Delgado}, {Di Matteo},
  {Diakite}, {Diener}, {Distefano}, {Dolding}, {Drazinos}, {Dur{\'a}n},
  {Edvardsson}, {Enke}, {Eriksson}, {Esquej}, {Eynard Bontemps}, {Fabre},
  {Fabrizio}, {Faigler}, {Falc{\~a}o}, {Farr{\`a}s Casas}, {Federici},
  {Fedorets}, {Fernique}, {Figueras}, {Filippi}, {Findeisen}, {Fonti},
  {Fraile}, {Fraser}, {Fr{\'e}zouls}, {Gai}, {Galleti}, {Garabato},
  {Garc{\'\i}a-Sedano}, {Garofalo}, {Garralda}, {Gavel}, {Gavras}, {Gerssen},
  {Geyer}, {Giacobbe}, {Gilmore}, {Girona}, {Giuffrida}, {Glass}, {Gomes},
  {Granvik}, {Gueguen}, {Guerrier}, {Guiraud}, {Guti{\'e}rrez-S{\'a}nchez},
  {Haigron}, {Hatzidimitriou}, {Hauser}, {Haywood}, {Heiter}, {Helmi}, {Heu},
  {Hilger}, {Hobbs}, {Hofmann}, {Holland}, {Huckle}, {Hypki}, {Icardi},
  {Jan{\ss}en}, {Jevardat de Fombelle}, {Jonker}, {Juh{\'a}sz}, {Julbe},
  {Karampelas}, {Kewley}, {Klar}, {Kochoska}, {Kohley}, {Kolenberg},
  {Kontizas}, {Kontizas}, {Koposov}, {Kordopatis}, {Kostrzewa-Rutkowska},
  {Koubsky}, {Lambert}, {Lanza}, {Lasne}, {Lavigne}, {Le Fustec}, {Le
  Poncin-Lafitte}, {Lebreton}, {Leccia}, {Leclerc}, {Lecoeur-Taibi},
  {Lenhardt}, {Leroux}, {Liao}, {Licata}, {Lindstr{\o}m}, {Lister}, {Livanou},
  {Lobel}, {L{\'o}pez}, {Managau}, {Mann}, {Mantelet}, {Marchal}, {Marchant},
  {Marconi}, {Marinoni}, {Marschalk{\'o}}, {Marshall}, {Martino}, {Marton},
  {Mary}, {Massari}, {Matijevi{\v{c}}}, {Mazeh}, {McMillan}, {Messina},
  {Michalik}, {Millar}, {Molina}, {Molinaro}, {Moln{\'a}r}, {Montegriffo},
  {Mor}, {Morbidelli}, {Morel}, {Morris}, {Mulone}, {Muraveva}, {Musella},
  {Nelemans}, {Nicastro}, {Noval}, {O'Mullane}, {Ord{\'e}novic},
  {Ord{\'o}{\~n}ez-Blanco}, {Osborne}, {Pagani}, {Pagano}, {Pailler},
  {Palacin}, {Palaversa}, {Panahi}, {Pawlak}, {Piersimoni}, {Pineau}, {Plachy},
  {Plum}, {Poggio}, {Poujoulet}, {Pr{\v{s}}a}, {Pulone}, {Racero}, {Ragaini},
  {Rambaux}, {Ramos-Lerate}, {Regibo}, {Reyl{\'e}}, {Riclet}, {Ripepi}, {Riva},
  {Rivard}, {Rixon}, {Roegiers}, {Roelens}, {Romero-G{\'o}mez}, {Rowell},
  {Royer}, {Ruiz-Dern}, {Sadowski}, {Sagrist{\`a} Sell{\'e}s}, {Sahlmann},
  {Salgado}, {Salguero}, {Sanna}, {Santana-Ros}, {Sarasso}, {Savietto},
  {Schultheis}, {Sciacca}, {Segol}, {Segovia}, {S{\'e}gransan}, {Shih},
  {Siltala}, {Silva}, {Smart}, {Smith}, {Solano}, {Solitro}, {Sordo}, {Soria
  Nieto}, {Souchay}, {Spagna}, {Spoto}, {Stampa}, {Steele},
  {Steidelm{\"u}ller}, {Stephenson}, {Stoev}, {Suess}, {Surdej}, {Szabados},
  {Szegedi-Elek}, {Tapiador}, {Taris}, {Tauran}, {Taylor}, {Teixeira},
  {Terrett}, {Teyssand ier}, {Thuillot}, {Titarenko}, {Torra Clotet}, {Turon},
  {Ulla}, {Utrilla}, {Uzzi}, {Vaillant}, {Valentini}, {Valette}, {van Elteren},
  {Van Hemelryck}, {van Leeuwen}, {Vaschetto}, {Vecchiato}, {Veljanoski},
  {Viala}, {Vicente}, {Vogt}, {von Essen}, {Voss}, {Votruba}, {Voutsinas},
  {Walmsley}, {Weiler}, {Wertz}, {Wevers}, {Wyrzykowski}, {Yoldas},
  {{\v{Z}}erjal}, {Ziaeepour}, {Zorec}, {Zschocke}, {Zucker}, {Zurbach}, \&
  {Zwitter}}]{gaiaDR2}
{Gaia Collaboration}, {Brown}, A.~G.~A., {Vallenari}, A., {et~al.} 2018, \aap,
  616, A1

\bibitem[{{Gaia Collaboration} {et~al.}(2020){Gaia Collaboration}, {Brown},
  {Vallenari}, {Prusti}, {de Bruijne}, {Babusiaux}, \& {Biermann}}]{gaiaEDR3}
{Gaia Collaboration}, {Brown}, A.~G.~A., {Vallenari}, A., {et~al.} 2020, arXiv
  e-prints, arXiv:2012.01533

\bibitem[{{Gao}(2020)}]{gao2020a}
{Gao}, X. 2020, \apj, 894, 48

\bibitem[{{Goodman} \& {Weare}(2010)}]{goo2020}
{Goodman}, J. \& {Weare}, J. 2010, Communications in Applied Mathematics and
  Computational Science, 5, 65

\bibitem[{{Grice} \& {Dawson}(1990)}]{gri1990}
{Grice}, N.~A. \& {Dawson}, D.~W. 1990, \pasp, 102, 881

\bibitem[{{Heggie} \& {Hut}(2003)}]{heggie_hut_2003}
{Heggie}, D. \& {Hut}, P. 2003, {The Gravitational Million-Body Problem: A
  Multidisciplinary Approach to Star Cluster Dynamics}

\bibitem[{{Hillenbrand} \& {Hartmann}(1998)}]{hill1998}
{Hillenbrand}, L.~A. \& {Hartmann}, L.~W. 1998, \apj, 492, 540

\bibitem[{{Hu} {et~al.}(2021){Hu}, {Zhang}, {Esamdin}, {Liu}, \&
  {Zeng}}]{hu2021}
{Hu}, Q., {Zhang}, Y., {Esamdin}, A., {Liu}, J., \& {Zeng}, X. 2021, \apj, 912,
  5

\bibitem[{{Hunt} \& {Reffert}(2021)}]{hunt21}
{Hunt}, E.~L. \& {Reffert}, S. 2021, \aap, 646, A104

\bibitem[{{Jerabkova} {et~al.}(2021){Jerabkova}, {Boffin}, {Beccari}, {de
  Marchi}, {de Bruijne}, \& {Prusti}}]{jer21}
{Jerabkova}, T., {Boffin}, H. M.~J., {Beccari}, G., {et~al.} 2021, \aap, 647,
  A137

\bibitem[{{Kharchenko} {et~al.}(2013){Kharchenko}, {Piskunov}, {Schilbach},
  {R{\"o}ser}, \& {Scholz}}]{khar2013}
{Kharchenko}, N.~V., {Piskunov}, A.~E., {Schilbach}, E., {R{\"o}ser}, S., \&
  {Scholz}, R.~D. 2013, \aap, 558, A53

\bibitem[{{King}(1962)}]{King62}
{King}, I. 1962, \aj, 67, 471

\bibitem[{{Kounkel} \& {Covey}(2019)}]{kounkel20}
{Kounkel}, M. \& {Covey}, K. 2019, \aj, 158, 122

\bibitem[{{Krone-Martins} \& {Moitinho}(2014)}]{kro14}
{Krone-Martins}, A. \& {Moitinho}, A. 2014, \aap, 561, A57

\bibitem[{{Kroupa}(1995)}]{kro1995}
{Kroupa}, P. 1995, \mnras, 277, 1522

\bibitem[{{Krumholz} {et~al.}(2019){Krumholz}, {McKee}, \&
  {Bland-Hawthorn}}]{kru19}
{Krumholz}, M.~R., {McKee}, C.~F., \& {Bland-Hawthorn}, J. 2019, \araa, 57, 227

\bibitem[{{K{\"u}pper} {et~al.}(2010){K{\"u}pper}, {Kroupa}, {Baumgardt}, \&
  {Heggie}}]{kup2010}
{K{\"u}pper}, A. H.~W., {Kroupa}, P., {Baumgardt}, H., \& {Heggie}, D.~C. 2010,
  \mnras, 407, 2241

\bibitem[{{K{\"u}pper} {et~al.}(2008){K{\"u}pper}, {MacLeod}, \&
  {Heggie}}]{kup2008}
{K{\"u}pper}, A. H.~W., {MacLeod}, A., \& {Heggie}, D.~C. 2008, \mnras, 387,
  1248

\bibitem[{{Lada} \& {Lada}(2003{\natexlab{a}})}]{lad03}
{Lada}, C.~J. \& {Lada}, E.~A. 2003{\natexlab{a}}, \araa, 41, 57

\bibitem[{{Lada} \& {Lada}(2003{\natexlab{b}})}]{lada2003}
{Lada}, C.~J. \& {Lada}, E.~A. 2003{\natexlab{b}}, \araa, 41, 57

\bibitem[{{Lamers} \& {Gieles}(2006)}]{lam2006}
{Lamers}, H.~J.~G.~L.~M. \& {Gieles}, M. 2006, \aap, 455, L17

\bibitem[{{Lindegren} {et~al.}(2021){Lindegren}, {Bastian}, {Biermann},
  {Bombrun}, {de Torres}, {Gerlach}, {Geyer}, {Hern{\'a}ndez}, {Hilger},
  {Hobbs}, {Klioner}, {Lammers}, {McMillan}, {Ramos-Lerate},
  {Steidelm{\"u}ller}, {Stephenson}, \& {van Leeuwen}}]{lin2021}
{Lindegren}, L., {Bastian}, U., {Biermann}, M., {et~al.} 2021, \aap, 649, A4

\bibitem[{{Liu} \& {Pang}(2019)}]{liu2019}
{Liu}, L. \& {Pang}, X. 2019, \apjs, 245, 32

\bibitem[{{Maschberger} {et~al.}(2010){Maschberger}, {Clarke}, {Bonnell}, \&
  {Kroupa}}]{mas2010}
{Maschberger}, T., {Clarke}, C.~J., {Bonnell}, I.~A., \& {Kroupa}, P. 2010,
  \mnras, 404, 1061

\bibitem[{{Mathieu}(1984)}]{mat1984}
{Mathieu}, R.~D. 1984, \apj, 284, 643

\bibitem[{McInnes {et~al.}(2017)McInnes, Healy, \& Astels}]{hdbscan_python}
McInnes, L., Healy, J., \& Astels, S. 2017, Journal of Open Source Software, 2,
  205

\bibitem[{{McMillan} {et~al.}(2007){McMillan}, {Vesperini}, \& {Portegies
  Zwart}}]{mcmi2007}
{McMillan}, S. L.~W., {Vesperini}, E., \& {Portegies Zwart}, S.~F. 2007, \apjl,
  655, L45

\bibitem[{{Meingast} \& {Alves}(2019)}]{mei2019}
{Meingast}, S. \& {Alves}, J. 2019, \aap, 621, L3

\bibitem[{{Meingast} {et~al.}(2021){Meingast}, {Alves}, \&
  {Rottensteiner}}]{mei21}
{Meingast}, S., {Alves}, J., \& {Rottensteiner}, A. 2021, \aap, 645, A84

\bibitem[{{Nilakshi} {et~al.}(2002){Nilakshi}, {Sagar}, {Pandey}, \&
  {Mohan}}]{nit2002}
{Nilakshi}, {Sagar}, R., {Pandey}, A.~K., \& {Mohan}, V. 2002, \aap, 383, 153

\bibitem[{{Nony} {et~al.}(2021){Nony}, {Robitaille}, {Motte}, {Gonzalez},
  {Joncour}, {Moraux}, {Men'shchikov}, {Didelon}, {Louvet}, {Buckner},
  {Schneider}, {Lumsden}, {Bontemps}, {Pouteau}, {Cunningham}, {Fiorellino},
  {Oudmaijer}, {Andr{\'e}}, \& {Thomasson}}]{nony2021}
{Nony}, T., {Robitaille}, J.~F., {Motte}, F., {et~al.} 2021, \aap, 645, A94

\bibitem[{{Olivares} {et~al.}(2018){Olivares}, {Moraux}, {Sarro}, {Bouy},
  {Berihuete}, {Barrado}, {Huelamo}, {Bertin}, \& {Bouvier}}]{oli18}
{Olivares}, J., {Moraux}, E., {Sarro}, L.~M., {et~al.} 2018, \aap, 612, A70

\bibitem[{{Oswalt} \& {Gilmore}(2013)}]{reviewOCsfriel}
{Oswalt}, T.~D. \& {Gilmore}, G. 2013, {Planets, Stars and Stellar Systems Vol.
  5}

\bibitem[{{Pang} {et~al.}(2021){Pang}, {Li}, {Yu}, {Tang}, {Dinnbier},
  {Kroupa}, {Pasquato}, \& {Kouwenhoven}}]{pang2021}
{Pang}, X., {Li}, Y., {Yu}, Z., {et~al.} 2021, \apj, 912, 162

\bibitem[{Pedregosa {et~al.}(2011)Pedregosa, Varoquaux, Gramfort, Michel,
  Thirion, Grisel, Blondel, Prettenhofer, Weiss, Dubourg, Vanderplas, Passos,
  Cournapeau, Brucher, Perrot, \& {{\'E}}douard Duchesnay}]{sklearn}
Pedregosa, F., Varoquaux, G., Gramfort, A., {et~al.} 2011, Journal of Machine
  Learning Research, 12, 2825

\bibitem[{{Piecka} \& {Paunzen}(2021)}]{pie2021}
{Piecka}, M. \& {Paunzen}, E. 2021, arXiv e-prints, arXiv:2107.07230

\bibitem[{{Piskunov} {et~al.}(2007){Piskunov}, {Schilbach}, {Kharchenko},
  {R{\"o}ser}, \& {Scholz}}]{pisk2007}
{Piskunov}, A.~E., {Schilbach}, E., {Kharchenko}, N.~V., {R{\"o}ser}, S., \&
  {Scholz}, R.~D. 2007, \aap, 468, 151

\bibitem[{{Plunkett} {et~al.}(2018){Plunkett}, {Fern{\'a}ndez-L{\'o}pez},
  {Arce}, {Busquet}, {Mardones}, \& {Dunham}}]{plu2018}
{Plunkett}, A.~L., {Fern{\'a}ndez-L{\'o}pez}, M., {Arce}, H.~G., {et~al.} 2018,
  \aap, 615, A9

\bibitem[{{Rom{\'a}n-Z{\'u}{\~n}iga} {et~al.}(2019){Rom{\'a}n-Z{\'u}{\~n}iga},
  {Alfaro}, {Palau}, {Hasenberger}, {Alves}, {Lombardi}, \&
  {S{\'a}nchez}}]{zun2019}
{Rom{\'a}n-Z{\'u}{\~n}iga}, C.~G., {Alfaro}, E., {Palau}, A., {et~al.} 2019,
  \mnras, 489, 4429

\bibitem[{{R{\"o}ser} \& {Schilbach}(2019)}]{roe_praesepe}
{R{\"o}ser}, S. \& {Schilbach}, E. 2019, \aap, 627, A4

\bibitem[{{R{\"o}ser} {et~al.}(2019){R{\"o}ser}, {Schilbach}, \&
  {Goldman}}]{roe2019}
{R{\"o}ser}, S., {Schilbach}, E., \& {Goldman}, B. 2019, \aap, 621, L2

\bibitem[{{Schilbach} {et~al.}(2006){Schilbach}, {Kharchenko}, {Piskunov},
  {R{\"o}ser}, \& {Scholz}}]{sch2006}
{Schilbach}, E., {Kharchenko}, N.~V., {Piskunov}, A.~E., {R{\"o}ser}, S., \&
  {Scholz}, R.~D. 2006, \aap, 456, 523

\bibitem[{{Sim} {et~al.}(2019){Sim}, {Lee}, {Ann}, \& {Kim}}]{sim19}
{Sim}, G., {Lee}, S.~H., {Ann}, H.~B., \& {Kim}, S. 2019, Journal of Korean
  Astronomical Society, 52, 145

\bibitem[{{Tang} {et~al.}(2019){Tang}, {Pang}, {Yuan}, {Chen}, {Hong},
  {Goldman}, {Just}, {Shukirgaliyev}, \& {Lin}}]{tang2019}
{Tang}, S.-Y., {Pang}, X., {Yuan}, Z., {et~al.} 2019, \apj, 877, 12

\bibitem[{{Taylor}(2005)}]{Taylor2005}
{Taylor}, M.~B. 2005, in Astronomical Society of the Pacific Conference Series,
  Vol. 347, Astronomical Data Analysis Software and Systems XIV, ed.
  P.~{Shopbell}, M.~{Britton}, \& R.~{Ebert}, 29

\bibitem[{{van de Ven} {et~al.}(2006){van de Ven}, {van den Bosch}, {Verolme},
  \& {de Zeeuw}}]{vdw06}
{van de Ven}, G., {van den Bosch}, R.~C.~E., {Verolme}, E.~K., \& {de Zeeuw},
  P.~T. 2006, \aap, 445, 513

\bibitem[{Virtanen {et~al.}(2020)Virtanen, Gommers, Oliphant, Haberland, Reddy,
  Cournapeau, Burovski, Peterson, Weckesser, Bright, {van der Walt}, Brett,
  Wilson, Millman, Mayorov, Nelson, Jones, Kern, Larson, Carey, Polat, Feng,
  Moore, {VanderPlas}, Laxalde, Perktold, Cimrman, Henriksen, Quintero, Harris,
  Archibald, Ribeiro, Pedregosa, {van Mulbregt}, \& {SciPy 1.0
  Contributors}}]{scipy}
Virtanen, P., Gommers, R., Oliphant, T.~E., {et~al.} 2020, Nature Methods, 17,
  261

\bibitem[{{Yeh} {et~al.}(2019){Yeh}, {Carraro}, {Montalto}, \&
  {Seleznev}}]{yeh2019}
{Yeh}, F.~C., {Carraro}, G., {Montalto}, M., \& {Seleznev}, A.~F. 2019, \aj,
  157, 115

\bibitem[{{Zhai} {et~al.}(2017){Zhai}, {Abt}, {Zhao}, \& {Li}}]{zhai2017}
{Zhai}, M., {Abt}, H., {Zhao}, G., \& {Li}, C. 2017, \aj, 153, 57

\bibitem[{{Zhang} {et~al.}(2020){Zhang}, {Tang}, {Chen}, {Pang}, \&
  {Liu}}]{zhang2020}
{Zhang}, Y., {Tang}, S.-Y., {Chen}, W.~P., {Pang}, X., \& {Liu}, J.~Z. 2020,
  \apj, 889, 99

\end{thebibliography}

\end{document}